\documentclass[
    superscriptaddress,
    amsmath,amssymb,
    aps,
    pra,
    onecolumn,
    ]{revtex4-2}
\usepackage{tcolorbox}
\usepackage{adjustbox}
\usepackage{xcolor}
\usepackage{graphicx} 
\usepackage{dcolumn} 
\usepackage{bm} 
\usepackage{braket} 
\usepackage{amsthm}
\usepackage{bbm}
\usepackage[normalem]{ulem}
\usepackage[ruled]{algorithm2e}
\usepackage{booktabs}
\usepackage{mathtools}
\usepackage{tabularray}
\newtheorem{theorem}{Theorem}

\newtheorem{lemma}{Lemma}

\newtheorem*{theorem-non}{Theorem}
\newtheorem*{lemma-non}{Lemma}
\newtheorem*{corollary-non}{Corollary}
\newtheorem*{proposition-non}{Proposition}
\usepackage{appendix} 
\usepackage[colorlinks]{hyperref}

\DeclareMathOperator{\Tr}{Tr}

\newcommand{\bx}{\bm{x}}
\newcommand{\bz}{\bm{z}}

\newcommand{\bxi}{\bm{x}^{(i)}}

\definecolor{lightgray}{rgb}{0.8, 0.8, 0.8}

\newcommand{\bomega}{\bm{\omega}}

\DeclareMathOperator{\RZ}{RZ}
\DeclareMathOperator{\RY}{RY}

\begin{document}

\title{Sample-efficient quantum error mitigation via classical learning surrogates} 

\author{Wei-You Liao}
\affiliation{Henan Key Laboratory of Quantum Information and Cryptography, Zhengzhou, Henan 450000, China}
\author{Ge Yan}
\affiliation{College of Computing and Data Science, Nanyang Technological University, Singapore, Singapore}
\author{Yujin Song}
\affiliation{College of Computing and Data Science, Nanyang Technological University, Singapore, Singapore}
\author{Tian-Ci Tian}
\author{Wei-Ming Zhu}
\author{De-Tao Jiang}
\affiliation{Henan Key Laboratory of Quantum Information and Cryptography, Zhengzhou, Henan 450000, China}
\author{Yuxuan Du}
\email{yuxuan.du@ntu.edu.sg}
\affiliation{College of Computing and Data Science, Nanyang Technological University, Singapore, Singapore}
\affiliation{School of Physical and Mathematical Science, Nanyang Technological University, Singapore, Singapore}
\author{He-Liang Huang}
\email{quanhhl@ustc.edu.cn}
\affiliation{Henan Key Laboratory of Quantum Information and Cryptography, Zhengzhou, Henan 450000, China}

\begin{abstract}
The pursuit of practical quantum utility on near-term quantum processors is critically challenged by their inherent noise. Quantum error mitigation (QEM) techniques are leading solutions to improve computation fidelity with relatively low qubit-overhead, while full-scale quantum error correction remains a distant goal. However, QEM techniques incur substantial measurement overheads, especially when applied to families of quantum circuits parameterized by classical inputs. Focusing on zero-noise extrapolation (ZNE), a widely adopted QEM technique, here we devise the surrogate-enabled ZNE (S-ZNE), which leverages classical learning surrogates to perform ZNE entirely on the classical side. Unlike conventional ZNE, whose measurement cost scales linearly with the number of circuits, S-ZNE requires only constant measurement overhead for an entire family of quantum circuits, offering superior scalability. Theoretical analysis indicates that S-ZNE achieves accuracy comparable to conventional ZNE in many practical scenarios, and numerical experiments on up to 100-qubit ground-state energy and quantum metrology tasks confirm its effectiveness. Our approach provides a template that can be effectively extended to other quantum error mitigation protocols, opening a promising path toward scalable error mitigation.
\end{abstract}

\maketitle

\section{INTRODUCTION}

\noindent Recent advances in quantum hardware fabrication have enabled demonstrations of computational capabilities approaching or even surpassing the classical frontier~\cite{aruteQuantumSupremacyUsing2019a,2021WuAdvantage,zhong2020quantum,madsen2022quantum,liu2025certified,huangSuperconductingQuantumComputing2020a}. However, the utility of these modern quantum processors is intrinsically limited by decoherence and operational errors, which degrade computational fidelity and obscure potential quantum utility~\cite{stilckfrancaLimitationsOptimizationAlgorithms2021,Gonz2022ErrorPropagation,morvanPhaseTransitionsRandom2024}. While full quantum error correction remains infeasible in the foreseeable future due to its immense resource overhead~\cite{gidney2025factor}, quantum error mitigation (QEM) has emerged as the critical strategy to suppress the noise-induced bias in expectation values on noisy hardware~\cite{Cai2023QEM,huangNeartermQuantumComputing2023a,2019ChaoPEC,zhangErrormitigatedQuantumGates2020}. On the one side, recent progress has validated QEM's indispensable role in many realistic applications, including ground-state energy estimation~\cite{kandalaErrorMitigationExtends2019,guoExperimentalQuantumComputational2024a} and digital quantum  simulation~\cite{miTimecrystallineEigenstateOrder2022,kim2023evidence,2025AlamFermionicdynamics}. On the other side, its principles are also poised to reduce the resource overhead of implementing quantum error correction protocols~\cite{2024KatabarwaQED,piveteau2021error,2021LostaglioMitigationAssisted,2022SuzukiQEMNISQtoFTQC,zhangDemonstratingQuantumError2025a,tsubouchiSymmetricCliffordTwirling2025,zhao2022realization, krinner2022realizing, google2023suppressing,googlequantumaiandcollaboratorsQuantumErrorCorrection2025,bluvstein2024logical}.

QEM has witnessed rapid theoretical and experimental advancements in recent years, yielding a diverse spectrum of strategies. Notable among them are zero-noise extrapolation (ZNE)~\cite{li2017efficient,2020Heidentityinsertions,cai2021multi,2023TsubouchiQEMEstimation,henaoAdaptiveQuantumError2023a,russo2025richard,kim2023evidence}, probabilistic error cancellation~\cite{temme2017error,2018EndoPracticalPEC,2021BravyiPEC,vandenbergProbabilisticErrorCancellation2023}, virtual distillation~\cite{Huggins2021VD,xuCircuitnoiseresilientVirtualDistillation2024,o2023purification,koczor2021exponential}, symmetry constraints \cite{PhysRevA.98.062339,McArdle2019symmtery}, learning-based approaches\cite{czarnik2021error,2025XuMLforQEM,liao2024machine,bennewitzNeuralErrorMitigation2022a,2021StrikisLearningBasedCDR}, and quantum subspace expansion~\cite{mcclean2020decoding,2020TakeshitaIncreasing}, among others. Owing to its conceptual simplicity and model-agnostic nature, ZNE has been successfully deployed on large-scale processors~\cite{kim2023evidence}, such as 127-qubit systems, to explore the computational utility of near-term quantum devices. Nevertheless, the practical adoption of QEM faces significant challenges, the most critical being the formidable sampling overhead required for its implementation~\cite{takagi2022fundamental,quek2024exponentially}. This sampling issue becomes particularly prohibitive for practical quantum tasks that involve families of circuits parameterized by classical inputs, as seen in variational quantum algorithms~\cite{2020McArdleQuantumcomputationalchemistry,cerezoVariationalQuantumAlgorithms2021,2022BhartiNISQ}, quantum many-body simulations~\cite{smithSimulatingQuantumManybody2019a,zhangDigitalQuantumSimulation2022,fauseweh2024quantum}, and quantum sensing~\cite{meyerVariationalToolboxQuantum2021, 2022Yangvariationalprinciple,2024ZengGeodesic,maclellanEndtoendVariationalQuantum2024} applications. In such scenarios, each distinct parameter value necessitates an independent and often costly execution of the full QEM protocol, leading to a potentially unsustainable resource burden. Consequently, a pivotal direction for current research is the development of efficient QEM frameworks that can mitigate errors across an entire family of quantum circuits with reduced sampling costs.

To address this bottleneck, we propose the surrogate-enabled ZNE (S-ZNE) framework. The core of S-ZNE is leveraging classical learning surrogates~\cite{duEfficientLearningLinear2025,Liao2025surrogate,du2025artificial} to decouple data acquisition from quantum execution. While it follows the standard ZNE procedure that extrapolates from amplified noise levels, the key difference is that the requisite noisy expectation values are predicted by the surrogate rather than measured directly. Consequently, after an initial training phase of the surrogates, S-ZNE mitigates errors for a class of parametrized circuits with different inputs purely on the classical side, without any additional measurement overhead. This constant-overhead property provides a fundamental scaling advantage over conventional ZNE, whose resource demands grow linearly with respect to the number of tasks.

The advantages of S-ZNE are established through both rigorous theoretical analysis and extensive numerical validation. Theoretically, we prove that under many practical scenarios, S-ZNE achieves an error scaling that is provably comparable to conventional ZNE. In addition, we conduct systematic experiments to validate S-ZNE's performance on two distinct applications: ground state energy estimation and quantum metrology task~\cite{leibfried2004toward} on up to 100 qubits. The achieved results indicate that S-ZNE achieves performance comparable to that of conventional ZNE while significantly reducing the sampling overhead. These results position S-ZNE as a powerful and practical tool for enhancing the capabilities of near-term quantum computing.

\begin{figure*}[tb]
    \centering\includegraphics[width=1\textwidth]{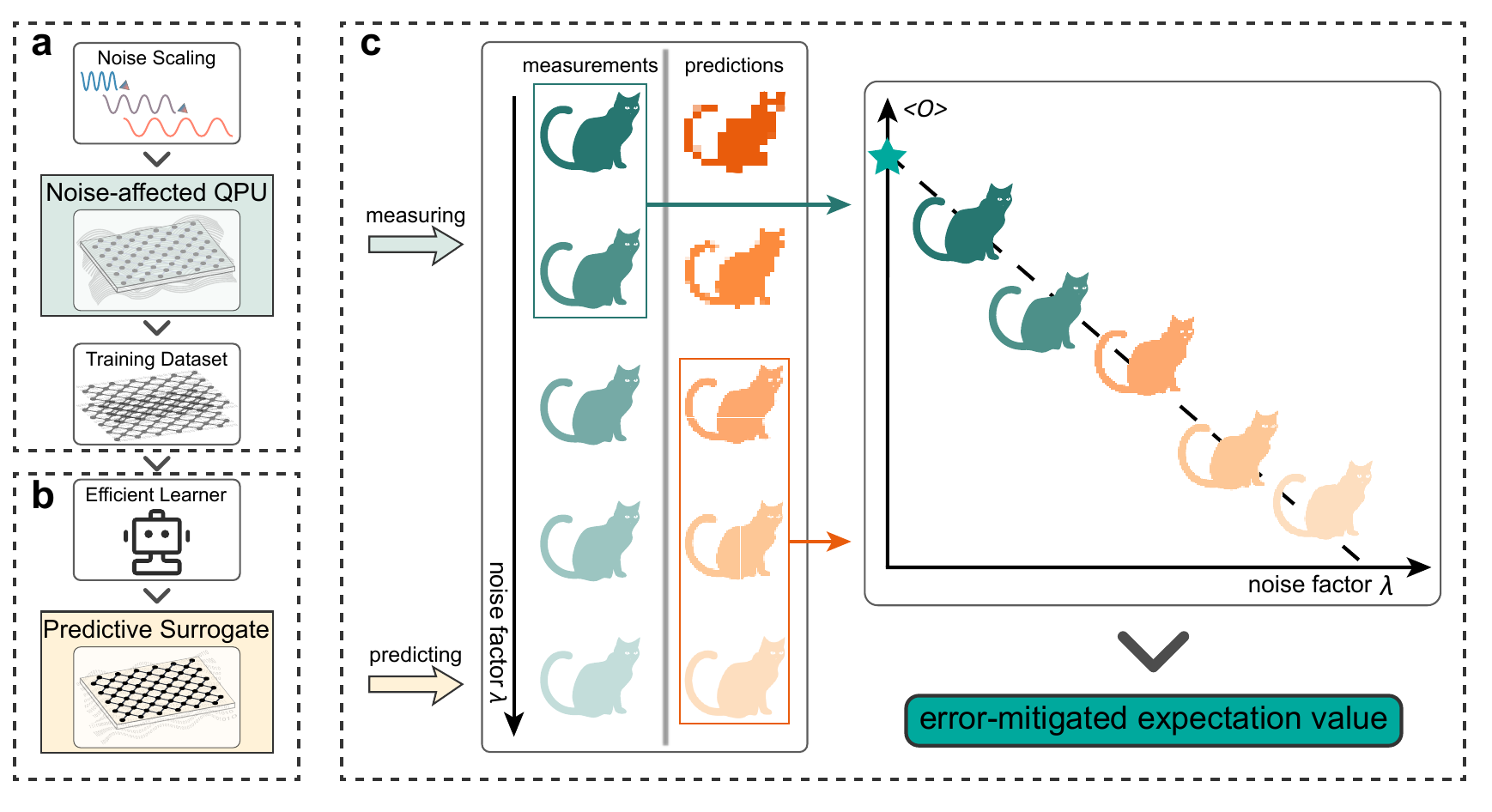}\vspace{-7pt}
    \caption{\small{\textbf{The scheme of surrogate-enabled zero noise extrapolation (S-ZNE).} 
    The S-ZNE framework towards a class of parametrized quantum circuits comprises three key stages. 
    \textbf{a.} {Data Collection}. Noise scaling (e.g., via unitary folding) is applied to generate the training dataset. At each noise level, different classical inputs are fed into the quantum circuit, followed by quantum sampling.   
    \textbf{b.} {Surrogate Modeling}. The constructed dataset trains classical learning surrogates to estimate expectation values under target noise conditions purely on the classical side. 
     \textbf{c.} {Surrogate-Enabled Extrapolation}: Given any new input, the optimized classical learning surrogates at all noise factors $\lambda$ can be used to predict the corresponding expectation values. In the main text, we use surrogate prediction for all noise levels, while a hybrid approach combining surrogate predictions and direct measurements from nosiy quantum processor (hybrid S-ZNE) is discussed in SI~E. These results extrapolate to $\lambda \to 0$, yielding an error-mitigated estimate with reduced quantum overhead.}}
\label{fig:scheme}
\end{figure*}

\section{Main RESULTS} 

\subsection{Problem setup} 
A crucial subroutine in a wide range of quantum algorithms involves the mean-value estimation for a family of parametrized quantum states controlled by classical inputs, i.e., $\{\rho(\bx)|\bx\in [0, 2\pi]^d\}$. Here $\rho(\bx)$ is generated by applying
a parametrized quantum circuit $U(\bx)$ to a fixed initial state $\rho_0$ with $\rho(\bx)=U(\bx)\rho_0U(\bx)^\dagger$. Concrete examples in this context include digital quantum simulation~\cite{lloyd1996universal}, variational quantum algorithms~\cite{cerezoVariationalQuantumAlgorithms2021}, and quantum metrology~\cite{leibfried2004toward}. Without loss of generality, $U(\bx)$
can always be decomposed into a set of Clifford gates (i.e., \{$H, S, \text{CNOT}\}$) and rotational gates along the Z-axis (i.e., $\RZ(\bx)$), which form a universal gate set~\cite{nielsen2010quantum}. Mathematically, we have
\begin{equation}\label{eq:unitary_ansatz}
    U(\bx)=\Big(\prod_{j=1}^d C_j\RZ(\bx_j)\Big)C_0,
\end{equation}
where $C_j$ refers to the unitary at the $j$-th layer formed by Clifford gates, and $d$ denotes the dimension of classical inputs $\bx$. The mean-value outcome of $\rho(\bx)$ for a given observable $O$ is denoted by $f(\bx, O)=\Tr(\rho(\bx) O)$. When deployed on noisy quantum processors, such parametrized states are often corrupted by a noise channel $\mathcal{N}_{\lambda}(\cdot)$ with $\lambda$ being the noise level. Accordingly, the experimentally accessible value is
\begin{equation}\label{eqn:main}
    f(\bx, O, \lambda)=\Tr\Big(\mathcal{N}_\lambda\big(\rho(\bx)\big) O\Big)
\end{equation}
instead of $f(\bx,O)$.
Throughout this study, we model the dominant noise as  Pauli channels, as any decoherent noise process can be transformed into Pauli noise through Pauli twirling~\cite{Cai2023QEM}.

Zero-noise extrapolation (ZNE) is a technique designed to estimate the noiseless mean-value outcome $f(\bx, O)$ by strategically leveraging the noisy estimations in Eq.~(\ref{eqn:main}). As shown in Fig.~\ref{fig:scheme}, the ZNE protocol involves systematically amplifying the noise channel $\mathcal{N}_\lambda$ with the level $\lambda$, e.g., via unitary folding~\cite{giurgica2020digital}. Denote $\mathcal{N}_0$ the identity channel, which leaves any quantum state unchanged. The ideal expectation value yields $f(\bx,O)\equiv f(\bx, O, \lambda=0)$. In practice, the noisy level $\lambda$ can only be amplified rather than reduced. Accordingly, ZNE obtains expectation values at a series of elevated effective noise levels $\{\lambda_j\}_{j=1}^u\in \mathbb{N}_+$ with $\lambda_j<\lambda_{j+1}$. After that, the desired noiseless result $f(\bx,O)$ is estimated through extrapolation. Specifically, let $g(\cdot)$ be the employed extrapolation function and $\hat{f}(\bx,O,\lambda_j)$ be an estimation of each noisy expectation  $f(\bx,O,\lambda_j)$ with $M$ snapshots. The estimation of the zero-noise limit, i.e., $f(\bx,O)$, yields
\begin{equation}\label{eqn:ZNE}
g(\bz_C(\bx))\equiv  g\left(\Big[\hat{f}(\bx, O, \lambda_1), \cdots, \hat{f}(\bx, O, \lambda_u)\Big]\right).
\end{equation}
The entire ZNE procedure is inherently measurement-intensive. That is, as each artificially amplified noise level $\lambda_j$ must be independently sampled, the total sampling cost grows linearly with the boost factor $u$. In addition, when applying ZNE to mitigate $f(\bx,O)$, the overhead further scales with the number of classical inputs $\bx$, rendering a substantial sampling overhead.

\subsection{Implementation of S-ZNE} To alleviate the substantial measurement overhead of ZNE when estimating $f(\bx,O)$ over varying inputs $\bx$, we propose the surrogate-ZNE (S-ZNE), a framework that leverages classical learning surrogates to predict $f(\bx, O, \lambda_j)$ and thereby circumvent repeated quantum hardware evaluations. For clarity, let us first recap the mechanism of classical learning surrogates, followed by presenting the implementation of S-ZNE.

The objective of classical learning surrogates~\cite{du2025artificial} is to effectively predict the expectation values of target observables across $\{\rho(\bx)\}$. An attractive feature of this approach is that once the dataset is collected from quantum hardware, both the training and inference processes are entirely classical. For example, the training dataset $\mathcal{T}=\{\bxi,y^{(i)}\}_{i=1}^n$ can be constructed by directly measuring the expectation values on quantum hardware, where $\bxi$ refers to the $i$-th training example and $y^{(i)}$ is the estimated expectation value of the observable $O$ for the (noisy) state $\rho(\bxi)$ with $T$ measurement shots. Once $\mathcal{T}$ is constructed, the regression-based classical learning surrogate can be effectively implemented, i.e., 
\begin{equation}\label{eq:surrogate}
    h(\bx,O)=\langle\boldsymbol{\Phi}_{\mathfrak{C}(\Lambda)}(\bx),\mathbf{w} \rangle,
\end{equation}
where $\boldsymbol{\Phi}_{\mathfrak{C}(\Lambda)}(\bx)$ denotes the feature vector composed of trigonometric basis functions with frequencies truncated at $\Lambda$. The weight vector $\mathbf{w}$ is obtained by solving the ridge regression problem with a regularization hyperparameter $\gamma > 0$, i.e.,
\begin{equation}
    \min_{\mathbf{w}} \Big\{\frac{1}{n} \sum_{i=1}^{n} \left( y^{(i)} - \langle \boldsymbol{\Phi}_{\mathfrak{C}(\Lambda)}(\boldsymbol{x}^{(i)}), \mathbf{w} \rangle \right)^2 + \gamma \| \mathbf{w} \|_2^2 \Big\}. \nonumber
\end{equation}
Refer to SI~A for implementation details.

An overview of S-ZNE is illustrated in Fig.~\ref{fig:scheme}. The philosophy of S-ZNE is to decouple noise extrapolation from repeated quantum executions by using classical surrogate models to predict circuit outcomes for new inputs. Mathematically, denote the learning surrogate trained on $f(\bx,O,\lambda_j)$ as $h(\bx,O,\lambda_j)$. For any new input $\bx'$, the surrogate outputs $h(\bx',O,\lambda_j)$ as the prediction of $f(\bx',O,\lambda_j)$ instead of $\hat{f}(\bx',O,\lambda_j)$ in Eq.~(\ref{eqn:ZNE}). As the prediction is conducted purely on the classical side, S-ZNE is able to substantially reduce the sampling overhead.  

The implementation of S-ZNE comprises three steps, which are data collection, surrogate modeling, and ZNE estimation. At the first stage, S-ZNE performs a one-time investment in quantum measurements to construct the training $\mathcal{T}(\lambda_j)=\{\bm{x}^{(i,j)}, y^{(i,j)}\}_{i=1}^{n_j}$ at each noise level $\lambda_j$, where $n_j$ denotes the number of training examples used in $h(\bx, O, \lambda_j)$ and $y^{(i,j)}$ refers to the mean-value estimation of $f(\bx, O, \lambda_j)$ with $T$ snapshots. Once $u$ training datasets are prepared, S-ZNE separately implement $u$ classical learning surrogates $\{h(\bx, O, \lambda_j)\}_{j=1}^u$ following Eq.~(\ref{eq:surrogate}). Last, S-ZNE employs the optimized $u$ classical learning surrogates to complete ZNE estimation purely on the classical side. According to Eq.~(\ref{eqn:ZNE}), the zero-noise limit  $f(\bx, O)$ for any new input $\bx$ is estimated by \[g(\bz_S(\bx))=g([h(\bx, O, \lambda_{1}),\cdots,h(\bx, O, \lambda_u)]).\]

We analyze the estimation error of S-ZNE relative to the conventional $\text{ZNE}$, as stated in the theorem below, whose proof is deferred to SI~B.
\begin{theorem}[Informal]\label{the:error_bound_comparison}
Following notations in Eqs.~(\ref{eq:unitary_ansatz})-(\ref{eq:surrogate}), let $L$ be the Lipschitz constant of $g(\cdot)$, and $\zeta^2$ be the intrinsic extrapolation error. Suppose that $\bm{x} \in [-R,R]^d$ is sampled from a distribution $\mathbb{D}$. When the conventional ZNE costs $M$ measurements to obtain $\hat{f}(\bx, O, \lambda_j)$, with probability at least $1-0.05u$, its average performance yields  
$\mathbb{E}_{\bm{x}}|f(\bm{x},O)-g(\bm{z}_C(\bm{x}))|^2 \le \zeta^2 + \mathcal{O}(L^2 u B^2/M)$. In addition, when $\Lambda>deq(1+R)$ and $n_j \geq \tilde{O}(B^2M (de/{\Lambda})^{4\Lambda})$,
with probability at least $1-u\delta$, the average performance of S-ZNE is upper bounded by $\mathbb{E}_{\bm{x}}|f(\bm{x},O)-g(\bm{z}_S(\bm{x}))|^2 \le \zeta^2 + \mathcal{O}(L^2 u B^2/M)$.
\end{theorem}

The achieved results deliver several implications. First, the performance of both conventional ZNEs and S-ZNE depends on $\zeta^2$. This error term stems from the choice of the extrapolation function $g(\cdot)$ and is independent of the approach used to estimate expectation values $f(\bx, O, \lambda_j)$ for the varied $j$. Moreover, the performance of both protocols is influenced by the Lipschitz constant $L$. As derived in SI~B, for the linear extrapolation, $L$ does not dominate the overall error, which scales approximately as $L\rightarrow 4/u$ as $u$ increases. Besides, the derived sample complexity $n_j$ reflects the efficiency of S-ZNE. In many realistic settings, the truncation parameter $\Lambda$ remains a small constant, e.g., noisy circuits with a large $q$ and quantum simulation tasks characterized by a small $R$. Under such conditions, S-ZNE is computationally efficient and can achieve performance comparable to that of conventional ZNE.

\noindent\textbf{Remark}. The proposed S-ZNE framework is highly flexible and can be adapted to a wide range of quantum tasks. For example, the regression-based surrogate can be substituted with other provably efficient learning models. In addition, S-ZNE can be extended to a hybrid structure, where classical learning surrogates are employed only at high noise levels to further balance accuracy and efficiency. Refer to SI~C for more details.

\begin{figure*}[t!]
    \centering\includegraphics[width=0.9\textwidth]{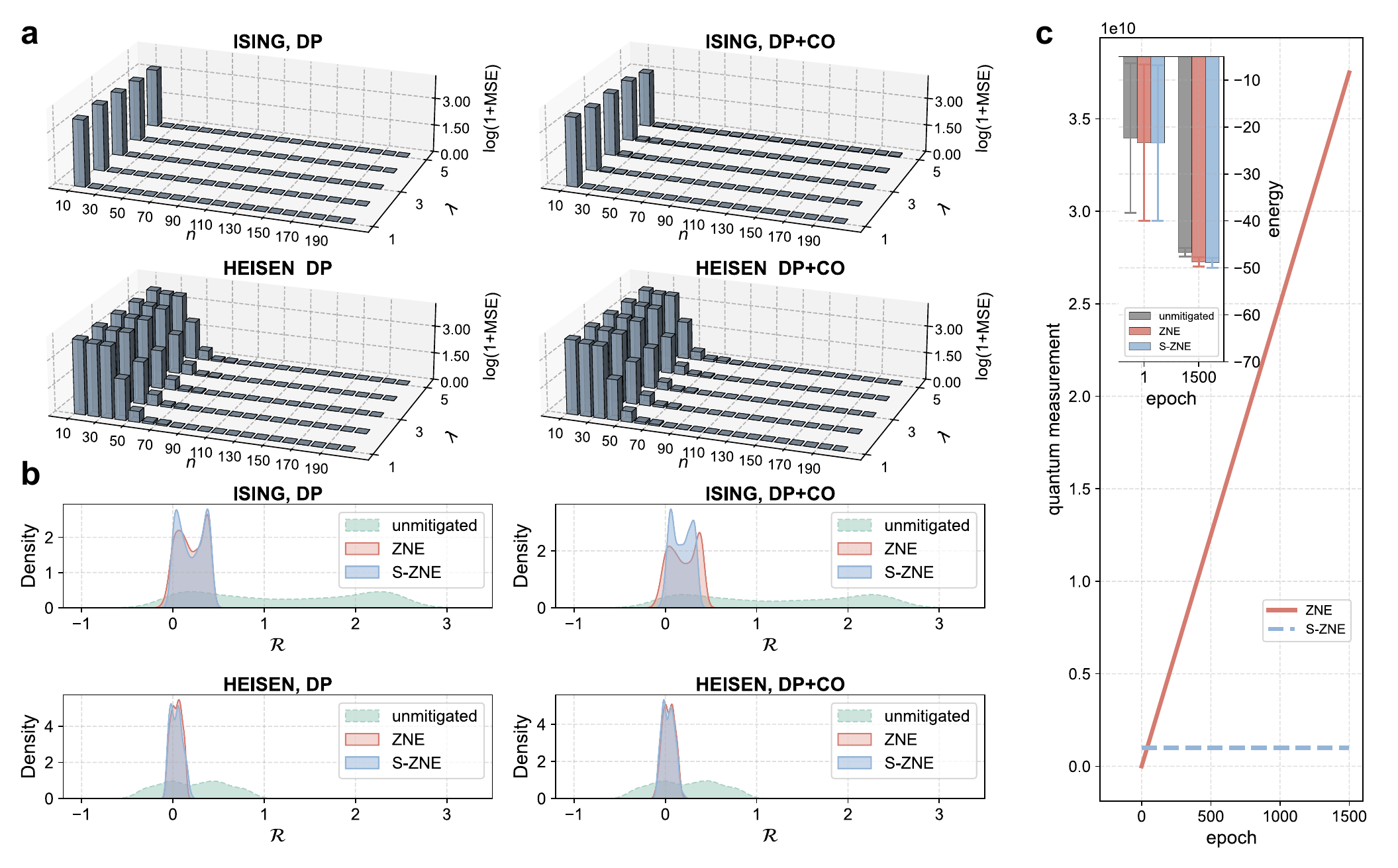}\vspace{-10pt}
    \caption{\small\textbf{Performance of S-ZNE in ground state energy estimation for quantum many-body systems.} 
    \textbf{a.} Surrogate model prediction accuracy under varying training set sizes ($n_j$) and noise levels ($\lambda_j$). Prediction error, quantified by $\log(1+\text{MSE})$, is evaluated on 1000 test samples for both transverse field Ising model (TFIM) and Heisenberg model (HM) under depolarizing (DP) and combined depolarizing+coherent (DP+CO) noise channels. 
    \textbf{b.} Distribution of residuals ($\mathcal{R}$) of unmitigated, S-ZNE, and conventional ZNE protocols, evaluated by the kernel density estimation over 1000 test configurations. 
    \textbf{c.} Sampling overhead analysis during the optimization. The main panel shows cumulative measurement costs for S-ZNE (constant overhead) versus ZNE (linear scaling with iteration). Inset displays estimated ground-state energies at epochs 1 and 1500 for TFIM (exact value: $-50.50$), demonstrating comparable convergence between S-ZNE and ZNE despite significantly reduced quantum resource requirements for S-ZNE.}
    \label{fig:fig2}
\end{figure*}

\section{Numerical Results}
Our first task is exploring the capabilities of S-ZNE in ground-state energy estimation using variational quantum algorithms (VQAs)~\cite{cerezoVariationalQuantumAlgorithms2021}. Given a specified ansatz $U(\bx)$, VQA estimates the ground-state energy of the Hamiltonian $\mathrm{H}$ by iteratively updating $\bx$ through gradient descent to minimize $f(\rho(\bx), \mathrm{H})$. Due to system noise, each iteration requires QEM to obtain reliable gradient information, incurring an increased sampling burden. To this end, S-ZNE offers an effective alternative to overcome this limitation.

To comprehend the effectiveness of S-ZNE, we consider two standard quantum many-body systems, i.e., the one-dimensional transverse field Ising models (TFIMs) and Heisenberg model (HMs) with open boundary conditions and $N=100$ qubits. Their mathematical forms are given by $\mathrm{H}_{\text{Ising}} = -\sum_{i} J Z_i Z_{i+1} - \sum_{i} h X_i$ and $\mathrm{H}_{\text{Heisen}} = \sum_{i} (J_x X_i X_{i+1} + J_y Y_i Y_{i+1}+ J_z Z_i Z_{i+1})$, respectively. For TFIM, we set  $J = 0.1$ and $h = 0.5$, while for HM, we set $J_x = 0.1$, $J_y = 0.5$ and $J_z=0$. For VQAs, we adopt the Hamiltonian variational ansatz~\cite{wecker2015progress}, where the implementation of  $U(\bx)$ depends on the explored Hamiltonian. The system noise is modeled by both the global depolarizing (DP) channel and the coherent (CO) channel. The depolarization rate is set as $1 - (1 - p_g)^{\lambda_j}$ with $p_g=0.05$. The coherent error is modeled by introducing perturbations on the input $\bx$, with the perturbation on each entry drawn uniformly from the range $[-0.01 \cdot \lambda_j,\, 0.01 \cdot \lambda_j]$. The extrapolation function $g(\cdot)$ is realized through a linear least-squares fit. Refer to SI~E for more details and additional results under diverse settings.

\begin{figure*}[tb!]
    \centering\includegraphics[width=0.8\textwidth]{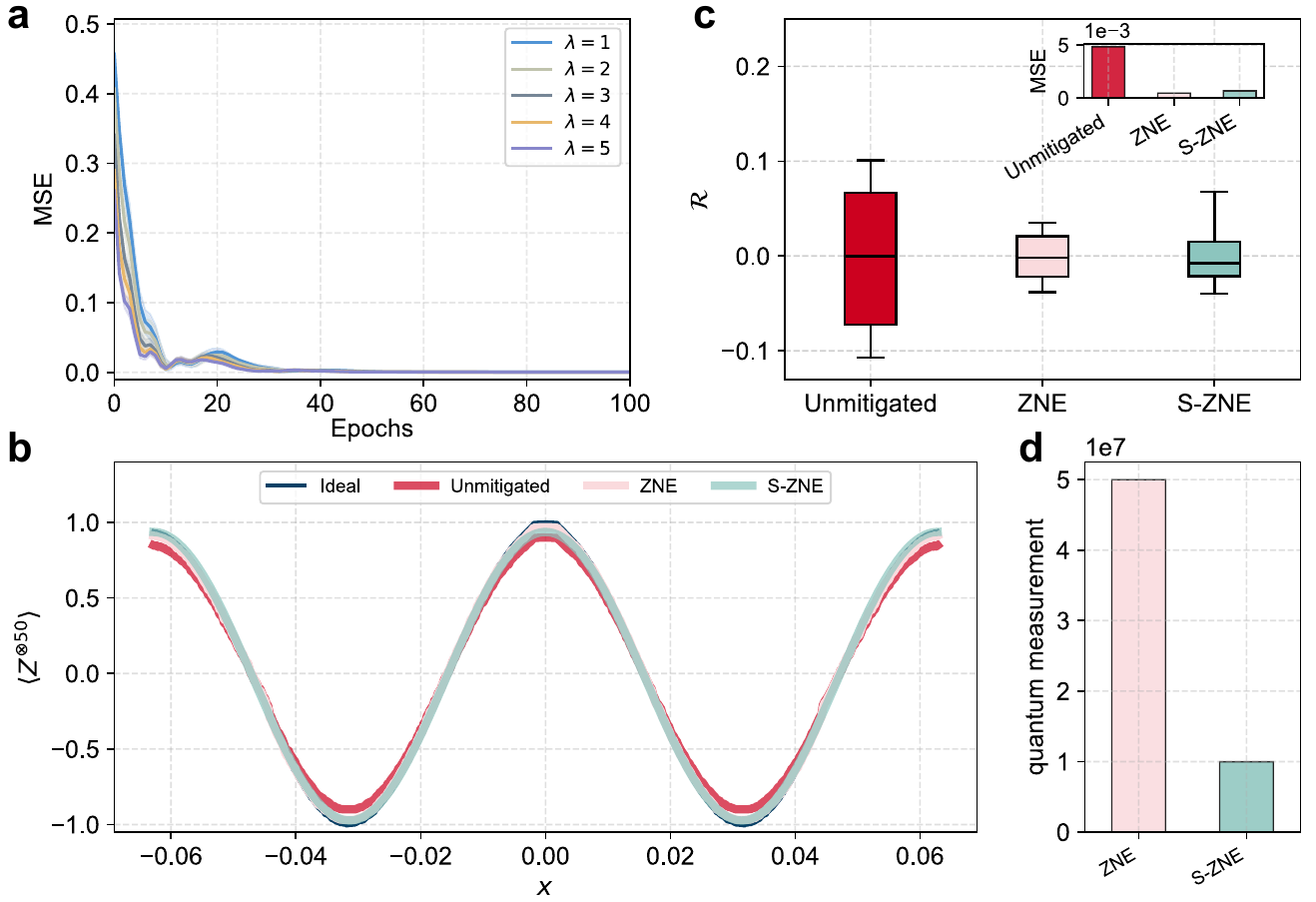}\vspace{-10pt}
    \caption{\small\textbf{Application of S-ZNE to GHZ-state quantum metrology.}
    \textbf{a.} Training dynamics of learning surrogates under varied noise levels $\lambda_i$. 
    \textbf{b.} Reconstructed phase estimation signals after quantum error mitigation. S-ZNE predictions and ZNE estimates both recover the ideal $2\pi/100$-periodic response.
    \textbf{c.} Residual analysis of error-mitigated expectation values. Both S-ZNE and ZNE maintain residuals close to the ideal $\cos(100x)$ signal, with an inset comparing overall MSE performance. 
    \textbf{d.} Sampling overhead of ZNE and S-ZNE. The y-axis refers to the cumulative number of measurements consumed by S-ZNE/ZNE for error mitigation.}
    \label{fig:ghz}
\end{figure*}

We begin by assessing the predictive accuracy of  classical surrogates $\{h(\bm{x}, \mathrm{H}, \lambda_j)\}$ with $\mathrm{H}\in \{\mathrm{H}_{\text{Ising}}, \mathrm{H}_{\text{Heisen}}\}$ with varying the number training examples $n_j$. The prediction performance is quantified by  $\log(1+\text{MSE})$ over a test dataset with $1000$ unseen inputs, where MSE refers to the mean-square error. As shown in Fig.~\ref{fig:fig2}\text{a}, for both DP and DP+CO noisy settings, the prediction error rapidly decreases with the increased $n_j$. When $n_j$ exceeds 100, the prediction error for both Hamiltonians is ignorable. In addition, the prediction error exhibits only a weak dependence on the noise level $\lambda_j$, slightly decreasing at higher noise levels. 

We next compare the error mitigation performance between S-ZNE and ZNE by evaluating their residuals, i.e.,  $\mathcal{R}_S = g(\bm{z}_S(\bm{x})) - f(\bx, \mathrm{H})$  and $\mathcal{R}_C = g(\bm{z}_C(\bm{x})) - f(\bx,\mathrm{H})$, respectively. For all surrogates used in S-ZNE, the number of training examples is fixed to be $n_j=200$. For  conventional ZNE,  $M=1 \times 10^6$ measurements are used to estimate each $\hat{f}(\bm{x}, \mathrm{H}, \lambda_j)$. The residual distributions, estimated by kernel density estimation (KDE) \cite{1986Density} over $1000$ test samples, are adopted as the metric. As shown in Fig.~\ref{fig:fig2}b, S-ZNE residuals cluster tightly around zero, achieving a distribution statistically comparable to that of conventional ZNE. This demonstrates that S-ZNE successfully mitigates errors to a similar level as ZNE under both DP and DP+CO noisy settings. 

We further analyze the consumed quantum resources for S-ZNE and ZNE. As shown in Fig.~\ref{fig:fig2}c, the inset shows a bar plot comparing the estimated ground-state energies of the TFIM at the 1st and $1500$-th iterations for S-ZNE, conventional ZNE, and the unmitigated cases, averaged over $10$ independent variational optimization runs. With the exact ground-state energy being $-50.50$, the results indicate that both S-ZNE and conventional ZNE provide comparable and significantly improved accuracy compared to the unmitigated case. In terms of resource efficiency, S-ZNE requires only a constant sampling overhead, consuming $1\times 10^9$ measurements throughout the entire optimization process. In contrast, the conventional ZNE necessitates $M \times u$ quantum measurements for every evaluation of $g(\bm{z}_C(\bm{x}))$, resulting in a linearly increasing sampling overhead that becomes dramatically larger than that of S-ZNE. The total cost of ZNE accumulates to $3.75\times 10^{10}$ measurements over the optimization course, which is 37.5 times that of S-ZNE.

\smallskip
Our second task applies S-ZNE to quantum metrology based on Ramsey interferometry~\cite{sackett2000experimental,leibfried2004toward}. The protocol uses a 100-qubit GHZ probe state,
$$
\ket{\text{GHZ}}_{100} = \frac{1}{\sqrt{2}}\bigl(\ket{0}^{\otimes 100} + \ket{1}^{\otimes 100}\bigr).
$$
The unknown phase $x$ is encoded via the global unitary $U(x) = \RZ(x)^{\otimes 100}$, which imprints a relative phase $\exp({\imath 100 x})$ between the two components of the GHZ state. The ideal signal is then given by  
\[
f(x, Z^{\otimes 100}) = \Tr\bigl(\rho(x)\, Z^{\otimes 100}\bigr) = \cos(100\,x),
\]
where $\rho(x) = U(x)\ket{\text{GHZ}}_{100}\!\bra{\text{GHZ}}\,U^\dagger(x)$. See SI~E for more details. In the subsequent experiments, the global depolarizing noise with the strength $p_g = 0.1$ is considered to damp the signal amplitude. 

We compare $\text{S-ZNE}$ with conventional ZNE using noisy data collected at $u=5$ noise amplification levels. For conventional ZNE, it estimates the noisy expectation values $\hat{f}(x, Z^{\otimes N}, \lambda_j)$ by using $M = 2 \times 10^4$ measurement shots per evaluation point. For $\text{S-ZNE}$, we train the learning surrogate in Eq.~(\ref{eq:surrogate}) with $n_j = 100$ training samples~($T=2 \times 10^4$) for $\forall j \in [5]$. As shown in Fig.~\ref{fig:ghz}a, the MSE of the surrogates consistently drops below $2 \times 10^{-4}$ within $100$ iterations across all noise factors, despite widely varying initial errors. This rapid and uniform convergence confirms that the surrogates reliably learn the noise-parametrized response even in this high-dimensional setting.

With stable surrogates in place, we assess phase estimation accuracy in terms of the residual $\mathcal{R}$. Fig.~\ref{fig:ghz}b displays the error-mitigated expectation values over a representative range of $x$, which is discretized into 500 points for evaluation. The $\text{S-ZNE}$ predictions, obtained by evaluating the trained surrogates at the inference stage, closely follow the ideal $\cos(100x)$ curve. Conventional ZNE, based on repeated quantum measurements, also recovers the signal accurately. Both methods preserve the $2\pi/100$ periodicity essential for Heisenberg-limited metrology. Quantitatively, as shown in Fig.~\ref{fig:ghz}c, conventional ZNE achieves an MSE of $4.83 \times 10^{-4}$, while S-ZNE attains an MSE of $6.79 \times 10^{-4}$---both significantly lower than the unmitigated MSE of $4.84 \times 10^{-3}$. 

Critically, after initial training, S-ZNE mitigates errors classically with constant overhead, fundamentally outperforming conventional ZNE's linear scaling with input number. In practical terms, conventional ZNE required $500 \times u \times M = 5\times10^7$ total measurement shots to evaluate all 500 phase values across 5 noise levels, while S-ZNE needed only $n_j \times u  \times T = 1 \times 10^7$ shots during training---yielding a 80\% reduction in quantum measurement cost. Crucially, this efficiency gap widens as the number of input points increases.

\section{Discussion}
In this study, we introduced S-ZNE,  a framework that substantially reduces the measurement overhead associated with conventional ZNE when applied to families of quantum circuits parameterized by
classical inputs. Experimental results on ground-state energy estimation and quantum metrology tasks indicate that S-ZNE attains comparable extrapolation accuracy with conventional ZNEs while significantly reducing the measurement cost. These findings establish S-ZNE as a practical tool for enhancing QEM capabilities on near-term quantum processors.

Several promising research avenues emerge from this work, spanning both QEM and classical learning surrogates. An important direction is extending classical learning surrogates to other QEM strategies beyond ZNE, such as virtual distillation~\cite{Huggins2021VD,koczor2021exponential}, probabilistic error cancellation~\cite{temme2017error}, and learning-based approaches~\cite{czarnik2021error}. Furthermore, integrating learning surrogates with experimental calibration procedures—such as real-time noise regime adaptation or online calibration—could enhance the model's practicality and robustness in experimental settings. Exploring the synergy between deep-learning-based surrogates~\cite{zhu2022flexible,melko2024language,wu2024variational,zhao2025rethink,wang2025aide} and these adaptive protocols may also open new avenues for developing more responsive and theoretically-grounded quantum error mitigation techniques.

\section*{ACKNOWLEDGMENTS}
Y.D. acknowledges the funding from SUG (025257-00001), NTU. H.-L.H. acknowledges support from the National Natural Science Foundation of China (Grant No. 12274464), and Natural Science Foundation of Henan (Grant No. 242300421049)

%\renewcommand{\appendixname}{SI}
%\renewcommand\thefigure{\thesection.\arabic{figure}}   
%\renewcommand\thetable{\thesection.\arabic{table}}
%\bibliography{citation}
 
%

\clearpage
\newpage
\onecolumngrid

\appendix

\tableofcontents

\section{Preliminary results}
\label{apd:correlated_construction}

This section presents the theoretical foundations and preliminary results underpinning the S-ZNE framework. In particular, SI~\ref{apd:PTM} provides the mathematical framework of Pauli transfer matrix and trigonometric expansion for both noiseless and noisy quantum circuits, SI~\ref{apd:surrogate} details the implementation of two classical learning surrogates, SI~\ref{append:A-3} summarizes the noise models employed in numerical simulations, SI~\ref{apd:sampling reduction} discusses the complementarity with other sampling reduction methods, and SI~\ref{app:extrap_method} elaborates on the extrapolation functions used in ZNE.

\subsection{Pauli transfer matrix and trigonometric expansion of quantum circuits} \label{apd:PTM}
\noindent\textbf{Pauli Transfer Matrix}.
Here we review how to use the Pauli-Liouville representation to formulate the quantum state and the observable. Denote $P_l$ as the $l$-th normalized Pauli operator with $P_l\in \frac{1}{\sqrt{2^N}}\{\mathbb{I},X,Y,Z\}^{\otimes N}$, which satisfies $\Tr(P_l P_k)=\delta_{lk}$. An arbitrary density matrix $\rho$ could be expanded by a set of normalized Pauli operators, i.e., 
\[
\rho=\sum_k c_k P_k, \quad \text{with} \quad \  c_k=\Tr(P_k\rho).
\]
We can denote $\rho$ as a $4^N$-dimension vector under the normalized Pauli bases, i.e.,
\[
|\rho\rangle\rangle=[c_1,\cdots,c_k,\cdots,c_{4^N}]^{\top}. \]
Given a circuit $U(\bm{x})$, its representation under Pauli bases is termed as Pauli transfer matrix (PTM) \cite{2015GreenbaumQuantumGateSetTomography}. The matrix element $[\mathfrak{U}(\bm{x})]_{lk}$ at the $l$-th row and $k$-th column yields:
\begin{equation}
[\mathfrak{U}(\bm{x})]_{lk}= \Tr(P_l U(\bm{x})P_k U(\bm{x})^\dagger)=\langle\langle P_j|\mathfrak{U}(\bm{x})|P_K\rangle\rangle, \nonumber
\end{equation}
where $\ket{\rho}\rangle$ denotes the quantum state under PTM representation. To be concrete, the PTM representation of $\RZ(\bm{x}_j)$ gates takes the form as
\begin{equation}
\RZ(\bm{x}_j)=\left(\begin{array}{cccc}
 1&0&0&0\\
0&\text{cos}(\bm{x}_j)&-\text{sin}(\bm{x}_j)&0\\
0&\text{sin}(\bm{x}_j)&\text{cos}(\bm{x}_j)&0\\
0&0&0&1
\end{array}\right)=\mathrm{D}_0 +\text{cos}(\bm{x}_j) \mathrm{D}_1 +\text{sin}(\bm{x}_j) \mathrm{D}_{-1} 
\end{equation}
where
$
\mathrm{D}_0=\Big(\begin{smallmatrix}
 1&0&0&0\\
0&0&0&0\\
0&0&0&0\\
0&0&0&1
\end{smallmatrix}$\Big), $\mathrm{D}_1=\Big(\begin{smallmatrix}
 0&0&0&0\\
0&1&0&0\\
0&0&1&0\\
0&0&0&0
\end{smallmatrix}\Big)$, and 
$\mathrm{D}_{-1}=\Big(\begin{smallmatrix}
 0&0&0&0\\
0&0&-1&0\\
0&1&0&0\\
0&0&0&0
\end{smallmatrix}\Big)$.

\medskip
\noindent\textbf{Trigonometric expansion of noiseless quantum circuits}.
Following the main text, let us consider an $N$-qubit quantum circuit in the form of $
U(\bm{x})=\Big(\prod_{j=1}^{d} C_j \RZ(\bm{x}_j)\Big) C_0$. When applied to an arbitrary $N$-qubit input state $\rho_0$, the generated state can be reformulated by the trigonometric expansion under the PTM form, which is
\begin{equation}
\rho(\bm{x})=U(\bm{x})\rho_0U(\bm{x})^\dagger=\sum_{\bomega} \Phi_{\boldsymbol{\omega}}(\bm{x})\langle\langle\rho_0|\mathfrak{U}_{\boldsymbol{\omega}}^\dagger. \nonumber
\end{equation} 
Here the notation $\Phi_{\boldsymbol{\omega}}(\bx)$ with $\boldsymbol{\omega} \in \{0,1,-1\}^d$ refers to the trigonometric monomial basis
\begin{equation}
\Phi_{\bomega}(\bx)=\prod_{j=1}^d \begin{cases}1 & \text { if } \bomega_j=0 \\ \cos \left(\bx_j\right) & \text { if } \bomega_j=1 \\ \sin \left(\bx_j\right) & \text { if } \bomega_j=-1\end{cases}. \nonumber
\end{equation}
In addition, the matrix $\mathfrak{U}_{\bomega}$ is a combination of permutation and masking matrices. The expectation value $\Tr(\rho(\bx)O)$ can also be expressed using the trigonometric monomial bases, i.e.,
\begin{equation}\label{eq:expectation value}
    f(\rho(\bm{x}),O)\equiv \Tr(\rho(\bm{x})O)=\sum_{\bomega} \Phi_{\bomega}(\bm{x})\langle\langle\rho_0|\mathfrak{U}_{\boldsymbol{\omega}}^\dagger|O\rangle \rangle.
\end{equation}

\medskip
\noindent\textbf{Trigonometric expansion in the noisy scenario}. Ref.~\cite{Liao2025surrogate} generalizes the above mathematical expression into the noisy scenario. Denote a single-qubit Pauli channel and a multi-qubit Pauli channel by $\mathcal{N}_P$ and $\mathcal{M}$, respectively. The parametrized ansatz in $U(\bx)$ under the Pauli channel can be represented as 
\begin{equation}
    \tilde{U}(\bx)=\prod_{l=1}^d\Big(\tilde{C}_l\widetilde{\text{RZ}}(\bx_l)\Big)\tilde{C}_0,
\end{equation}
where $\widetilde{\text{RZ}}(\bx_l)=\mathcal{N}_{P}\circ\text{RZ}(\bx_l)$ and $\tilde{C}_l=\mathcal{M}_l\circ C_l$ with $M_l$ being a multi-qubit Pauli channel applied to the $l$-th Clifford operation $C_l$.

As with the noiseless case, this noisy unitary can be effectively converted into the PTM representation. This is because under PTM, both single-qubit and multi-qubit Pauli channels can be rewritten as fixed diagonal matrices. More specifically, recall the definition of a single-qubit Pauli channel, i.e., 
\begin{equation}\label{append:eqn:Pauli-channel}
\mathcal{N}_P[\rho]=(1-p_X-p_Y-p_Z)\rho+p_X X\rho X +p_Y Y \rho Y +p_Z Z \rho Z,
\end{equation}
where $p_X$, $p_Y$, and $p_Z$ denote the  Pauli error rates along $X$, $Y$, and $Z$ axes. Under the PTM representation, the single-qubit Pauli channel transforms into a diagonal matrix with \begin{equation}
\mathrm{N}= \text{diag} (1,q_X, q_Y,q_Z),
\end{equation}
where $q_X=1-2(p_Z+p_Y)$, $q_Y=1-2(p_Z+p_X)$, and $q_Z=1-2(p_X+p_Y)$. Similarly, an arbitrary multi-qubit Pauli channel $\mathcal{M}$ can be rewritten as a diagonal matrix under PTM representation.

Following the previous noiseless result, the noisy expectation value of (\ref{eq:expectation value}) can still be expanded into a set of trigonometric monomial bases, i.e.,
\begin{equation}
   f(\tilde{\rho}(\boldsymbol{x}), O) \equiv \operatorname{Tr}(\tilde{\rho}(\boldsymbol{x}) O)= \sum_{\boldsymbol{\omega}} \Phi_{\boldsymbol{\omega}}(\boldsymbol{x})\langle\langle O| \tilde{\mathfrak{U}}_{\boldsymbol{\omega}}|\rho_0\rangle\rangle.
\end{equation}
Compared to the noiseless case, the only difference is that $\{\tilde{\mathfrak{U}}_{\boldsymbol{\omega}}\}$ depends on the noisy rate of Pauli channels.

\subsection{Classical learning surrogates}\label{apd:surrogate}
In this subsection, we provide implementation details of two classical learning surrogates employed in S-ZNE, which are kernel-based and regression-based surrogates. Without loss of generality, we focus on elucidating the implementation of both classical learning surrogates when predicting the noisy expectation value at the $j$-th level, i.e., $f(\bx, O, \lambda_j)$.

\smallskip
\noindent\textbf{Kernel-based surrogate $h_{\mathsf{cs}}$}. The kernel-based learning surrogate is designed for circuits containing independent parameters $\bx$ and supporting varied observables with bounded locality. At the $j$-th noise level, when the unitary folding method is adopted, the explicit form of the circuit implementation under the Pauli channel is
\begin{equation}
    \tilde{U}(\bx; \lambda_j) = \prod_{k=1}^{\lambda_j} \left( \prod_{l=1}^d\Big(\tilde{C}_l\widetilde{\text{RZ}}(\bx_{l,k})\Big)\tilde{C}_0 \right) . \label{append:eqn:unit-folding}
\end{equation}
In other words, we repeat the implementation of $U(\bx)$ with $\lambda_j$ times. As a result, the total number of classical controls in the circuit increases to  $\lambda_j \times d$.   

During the data collection phase, the classical input $\bm{x}$ is randomly and uniformly sampled from $ [-\pi,\pi]^{\lambda_j \times d}$, and the Pauli-based classical shadows~\cite{huangPredictingManyProperties2020a} are used to acquire the classical representations of the resulting state prepared by $\tilde{U}(\bx; \lambda_j)$. The collected shadow state is denoted by $\tilde{\rho}_T(\bx)$ with $T$ being the number of snapshots. In this way, we can construct the training dataset $\mathcal{T}(\lambda_j)=\{\bx^{(i,j)}, \tilde{\rho}_T(\bx^{(i,j)})\}_{i=1}^{n_j}$ with $n_j$ training examples. 

Given access to the established training dataset $\mathcal{T}(\lambda_j)$, the kernel-based learning surrogate with respect to the observable $O$ yields
\begin{equation}\label{append:eqn:h-cs}
h_{\mathsf{cs}}(\bm{x},O, \lambda_j) = \frac{1}{n_j} \sum_{i=1}^{n_j} \kappa_{\Lambda}(\bm{x}, \bm{x}^{(i)}) \Tr(\tilde{\rho}_T(\bm{x}^{(i)})O),
\end{equation}
where $\kappa_{\Lambda}(\bm{x}, \bm{x}^{(i)}) =\sum_{\bomega \in \mathfrak{C}(\Lambda)} 2^{\|\bm{\omega}\|_0} \Phi_{\bm{\omega}}(\bm{x}) \Phi_{\bm{\omega}}(\bm{x}^{(i)})$ is the truncated trigonometric monomial kernel and the truncated frequency set is $\mathfrak{C}(\Lambda)=\{\bomega \in \{0,\pm1\}^d \mid \|\bomega\|_0 \le \Lambda\}$.  The feature map $\Phi_{\bm{\omega}}(\bm{x})$ is defined as:
\begin{equation}
\Phi_{\bm{\omega}}(\bm{x}) = \prod_{l=1}^d 
\begin{cases} 
1 & \text{if } \omega_l = 0, \\
\cos(x_l) & \text{if } \omega_l = 1, \\
\sin(x_l) & \text{if } \omega_l = -1.
\end{cases}
\end{equation}

\smallskip
The prediction performance of the kernel-based learning surrogate is warranted by the following lemma.
\begin{lemma}[{\cite[Adapted from Theorem~1]{Liao2025surrogate}}]\label{lemma:surrogate_error}
    Assume $\mathbb{E}_{\bm{x}\sim\mathrm{Unif}[-\pi,\pi]^d}\|\nabla_{\bm{x}}\Tr(\tilde{\rho}(\bm{x},\lambda_j)O)\|_2^2 \le C$. Let $O=\sum_i O_i$ be a $K$-local observable with $\sum_i \|O_i\|_{\infty} \le B$.  Consider a quantum circuit affected by a Pauli noise channel $\mathcal{N}_P(p_X,p_Y,p_Z)$, characterized by $p=\min\{p_X,p_Y\}$ and $p_Z$. Let $h_{\mathsf{cs}}(\bm{x}, O, \lambda_j)$ be the learning surrogate in Eq.~(\ref{append:eqn:h-cs}) trained on $n_j$ samples  with $\Lambda = \min\{\Lambda_C, \Lambda_p\}$, where $\Lambda_C = 4C/\epsilon_j$ and $\Lambda_p = \frac{1}{2(p+p_Z)} \log(2B/\sqrt{\epsilon_j})$. 
    
When the number of training examples satisfies  
\begin{equation}
        n_j \ge \left|\mathfrak{C}(\Lambda)\right| \frac{2 B^2 9^K}{\epsilon_j}\log\left(\frac{2 \cdot \left|\mathfrak{C}(\Lambda)\right|}{\delta}\right), \label{eq:N_requirement_lemma} \end{equation}
with probability at least $1-\delta$, the average prediction error of the kernel-based learning surrogate is bounded by
    \begin{equation}
        \mathbb{E}_{\bm{x}\sim\mathrm{Unif}[-\pi,\pi]^{d\times \lambda_j}}|h_{\mathsf{cs}}(\bm{x},O,\lambda_j) - f(\bm{x},O,\lambda_j)|^2 \le \epsilon_j.
    \end{equation}
\end{lemma}

We remark that ZNE with unitary folding requires the classical input to be correlated among $\lambda_j$ blocks. However, the performance guarantee of kernel-based learning surrogate rests on the independence of different entries. As such, when kernel-based learning surrogates are employed, the error bound of S-ZNE additionally depends on the domain generalization capability during inference. To this end, we conduct systematic numerical simulations to validate the capabilities of $h_{\mathsf{cs}}$ in S-ZNE. Refer to SI~\ref{app:additional_results} for more details.

\smallskip
\noindent\textbf{Regression-based surrogate $h_{\mathsf{qs}}$}. The regression-based surrogate applies to the scenario in which the classical controls are correlated. More specifically, the explored noisy quantum circuit at the $\lambda_j$-th level takes the form of 
\begin{equation}
\tilde{U}(\bx; \lambda_j) = \prod_{k=1}^{\lambda_j} \left( \prod_{l=1}^d\Big(\tilde{C}_l\widetilde{\text{RZ}}(\bx_{l})\Big)\tilde{C}_0 \right),
\end{equation}
where the $l$-th entry $\bx_l$ in $\bx$ repeats across $\lambda_j$ blocks. This formalism aligns with ZNE with unitary folding, as classical input $\bx$ is correlated in the circuit.   

The training dataset for the regression-based learning surrogate is defined as $\mathcal{T}(\lambda_j)=\{\bm{x}^{(i,j)}, y^{(i,j)}\}_{i=1}^{n_j}$. Here the input $\bx^{(i,j)}$ is sampled from an arbitrary distribution $\mathbb{D}$ within a bounded interval $[-R,R]^d$, and $y^{(i)}$ is the estimated mean value of the observable $O$ with $T$ shots. The mathematical expression of the regression-based surrogate is

\begin{equation}\label{append:eqn:h-qs}
h_{\mathsf{qs}}(\bm{x}, O, \lambda_j;\mathbf{w_j})=\langle\boldsymbol{\Phi}_{\mathfrak{C}(\Lambda)}(\bm{x}),\mathbf{w}_j \rangle, 
\end{equation}
where the high frequency terms with  $||\bm{\omega}||_0 > \Lambda$ is truncated. The weight $\mathbf{w}_j$ is obtained by solving the following ridge regression optimization problem, i.e.,
\begin{equation}
\min_{\mathbf{w}_j} \left\{\frac{1}{n_j} \sum_{i=1}^{n_j} \left( y^{(i,j)} - \langle \boldsymbol{\Phi}_{\mathfrak{C}(\Lambda)}(\boldsymbol{x}^{(i,j)}), \mathbf{w}_j \rangle \right)^2 + \gamma \| \mathbf{w}_j \|_2^2 \right\},
\end{equation}
where $\gamma > 0$ is a regularization hyperparameter. 
To explicitly capture parameter correlations, we partition the $d$-dimensional parameter vector $\bm{x}$ into $S$ groups ${\mathfrak{x}_1, \dots, \mathfrak{x}_S}$, where parameters in each group are identical. For each frequency vector $\bm{\omega} \in \mathfrak{C}(\Lambda)$, let $\omega_{s,k}$ denote the component of $\bm{\omega}$ corresponding to the $k$-th parameter in group $s$. The feature map is then defined as:
\begin{equation}
\Phi_{\bm{\omega}}(\bm{x}) = \prod_{s=1}^S \left[ \cos(\mathfrak{x}_s)^{N_s^+(\bm{\omega})} \cdot \sin(\mathfrak{x}_s)^{N_s^-(\bm{\omega})} \right],
\end{equation}
where $N_s^+(\bm{\omega}) = \sum_{k=1}^{d_s} \mathbbm{1}\{\omega_{s,k} = 1\}$ and $N_s^-(\bm{\omega}) = \sum_{k=1}^{d_s} \mathbbm{1}\{\omega_{s,k} = -1\}$ count the occurrences of cosine and sine terms in group $s$, respectively, with $d_s$ being the number of parameters in group $s$.

The performance error of the regression-based learning surrogate is provided in the following lemma.
\begin{lemma}[{\cite[Adapted from Theorem~2]{Liao2025surrogate}}] \label{lemma:surrogate_error_correlated}
 Following notations in Eq.~(\ref{append:eqn:h-qs}), let $q=1-2(p+p_Z)$ with $p=\min\{p_X,p_Y\}$,  $\Lambda$ be the threshold of the truncated frequency set $\mathfrak{C}(\Lambda)$, and $\epsilon_l$ be the maximal estimation error of $\{y^{(i,j)}\}$ in $\mathcal{T}(\lambda_j)$, namely, $\max_{i\in[n]}| y^{(i,j)}- f(\bx^{(i,j)})|\leq\epsilon_l$. Assume $q(1+R)<1/e$. Define $\epsilon=16B^2(deq(1+R)/\Lambda)^{2\Lambda}$. When the following conditions are satisfied: (i) $\epsilon_l\le\sqrt{\epsilon}/4$, (ii) the frequency is truncated to $\Lambda>deq(1+R)$, (iii) the number of training examples satisfies
    \begin{equation}
    n_j=\left(\frac{1}{q(1+R)}\right)^{4\Lambda}\cdot\frac{\log(1/\delta)}{9},\end{equation}
    the predictive surrogate $h_{\mathsf{qs}}(\bx)$ achieves with probability at least $1-\delta$,
    \begin{equation}
        \mathbb{E}_{\bx\sim\mathbb{D}}|h_{\mathsf{qs}}(\bx, O, \lambda_j)-f(\bm{x},O,\lambda_j)|^2\le\epsilon_j.
    \end{equation}
\end{lemma}

\noindent\textbf{Remark}.  Throughout the remainder of this study, we sometimes use $h_{\mathsf{qs}}$ and $h_{\mathsf{cs}}$ to specify which surrogate model is being referenced, so as to avoid confusion.

\subsection{A summary of the employed noisy models}\label{append:A-3}
This subsection details the standard noise models employed in numerical simulations, which include depolarization noise, thermal noise, and coherent noise.

\smallskip
\noindent\textbf{Depolarizing noise.}
We consider both \textit{local} and \textit{global} depolarizing noisy channels, both of which fall within the class of Pauli channels.

For an $N$-qubit system, the \textit{local} depolarizing channel acts independently on each qubit. It is defined as the tensor product of single-qubit depolarizing channels applied to each qubit. The single-qubit depolarizing channel is described by the Kraus operators
\begin{align}
K_0 &= \sqrt{1 - \tfrac{3p_d}{4}} \, \mathbb{I}, \quad
K_1 = \sqrt{\tfrac{p_d}{4}} \, X, \quad
K_2 = \sqrt{\tfrac{p_d}{4}} \, Y, \quad
K_3 = \sqrt{\tfrac{p_d}{4}} \, Z,
\end{align}
where $p_d \in [0,1]$ is the depolarizing rate per qubit. The corresponding map for one qubit is $\mathcal{E}_d(\rho) = (1-p_d)\rho + p_d \frac{\mathbb{I}}{2}$. For $N$ qubits, the local depolarizing channel is $\mathcal{E}_d^{\otimes N}$. 

The \textit{global} depolarizing channel acts collectively on all $N$ qubits. Its CPTP map is given by
\begin{align}
K_0 &= \sqrt{1 - p_g + \frac{p_g}{4^N}} \, \mathbb{I}^{\otimes N}, \\
K_\alpha &= \sqrt{\frac{p_g}{4^N}} \, P_\alpha, \quad \alpha = 1, 2, \dots, 4^N - 1,
\end{align}
where $\{P_\alpha\}$ denotes the set of all non-identity $N$-qubit Pauli operators, and $p_g \in [0,1]$ is the global depolarizing rate. This channel can be equivalently expressed as $\mathcal{N}_g(\rho) = (1-p_g)\rho + p_g (\mathbb{I}/{2})^{\otimes N}$.

\medskip
\noindent\textbf{Thermal noise}. Thermal relaxation noise combines energy dissipation ($T_1$ decay) and pure dephasing ($T_2$ decoherence). Let $t_g$ denote the gate duration. We define the reset probability and dephasing probability as $p_r = 1 - e^{-t_g/T_1}$ and $p_z = \tfrac{1}{2}\bigl(1 - e^{-t_g/T_\phi}\bigr)$, respectively. The pure dephasing time satisfies $T_\phi^{-1} = T_2^{-1} - (2T_1)^{-1}$. The following describes the single-qubit thermal relaxation channel, which is applied independently to each qubit in multi-qubit simulations.

\smallskip
\noindent\textit{Case $ T_2 \leq T_1 $}. The single-qubit thermal noisy channel is implemented via six Kraus operators
\begin{align}
& K_0 = \sqrt{1 - p_z - p_r} \, \mathbb{I}, \quad K_1 = \sqrt{p_z} \, Z, \quad
K_2 = \sqrt{p_r(1-p_e)} \, |0\rangle\langle 0|, \nonumber \\
& K_3 = \sqrt{p_r(1-p_e)} \, |0\rangle\langle 1|, \quad K_4 = \sqrt{p_r p_e} \, |1\rangle\langle 0|, \quad
K_5 = \sqrt{p_r p_e} \, |1\rangle\langle 1|,
\end{align}
where $ p_e \in [0,1] $ is the excited state population determined by the thermal equilibrium.

\smallskip
\noindent\textit{Case $ T_1 < T_2 $}. The single-qubit thermal noisy channel is implemented via the Choi matrix:
\[
\Lambda = \begin{bmatrix}
1 - p_e p_r & 0 & 0 & e^{-t_g/T_2} \\
0 & p_e p_r & 0 & 0 \\
0 & 0 & (1-p_e)p_r & 0 \\
e^{-t_g/T_2} & 0 & 0 & 1 - (1-p_e)p_r
\end{bmatrix}.
\]
This thermal noise model is specifically used in the experiments reported in SI~\ref{append:hybrid}.

\medskip
\noindent\textbf{Coherent noise}. We model coherent errors as small, stochastic miscalibrations of rotation angles in single-qubit rotation gates. For an ideal rotational gate $\mathrm{R}_P(x) = e^{-ix P/2}$ with $P \in \{X, Y, Z\}$, its noisy implementation is given by
\begin{equation}
    \mathrm{R}_P(x) \;\mapsto\; e^{-i(x + \theta_P)P/2},
\end{equation}
where the angular offset $\theta_P$ for each gate is drawn independently from a uniform distribution. This model captures both over-/under-rotation and axis misalignment effects.

\subsection{Complementary with other sampling reduction methods}
\label{apd:sampling reduction}

Prior studies relevant to S-ZNE can be broadly classified into two categories: (i) approaches that aim to reduce the sampling overhead of zero-noise extrapolation (ZNE) itself, and (ii) methods that mitigate sampling costs in specific quantum tasks, especially for variational quantum algorithms (VQAs). We discuss each in turn.

\smallskip
\noindent \textbf{ZNE variants}. To the best of our knowledge, only one study has explicitly sought to reduce the measurement overhead of conventional ZNE. In particular, Liao \textit{et al.}~\cite{liao2024machine} introduced random forest ZNE (RF-ZNE), which trains a random forest model to predict the noiseless expectation value directly from circuit descriptors and noisy outcomes, thereby bypassing explicit noise scaling and extrapolation at inference time. A key distinction between S-ZNE and RF-ZNE lies in their theoretical grounding: S-ZNE retains the rigorous physical interpretability of extrapolation under controlled noise scaling, whereas RF-ZNE operates as a data-driven surrogate without explicit noise modeling.

\smallskip
\noindent \textbf{Task-specific sampling overhead reduction}. While general-purpose strategies for reducing ZNE’s measurement cost remain scarce, substantial progress has been made in curbing sampling overhead in VQAs, whose optimization typically demands extensive quantum measurements due to non-convex loss landscapes~\cite{bittel2021training} and the forbidding of back-propagation~\cite{2023Abbasquantumbackpropagation}. To date, three principal lines of research have emerged to address this challenge: smart initialization, intelligent optimizers, and measurement grouping.

\noindent \textit{Smart initialization techniques}. Smart initialization techniques aim to reduce the number of quantum measurements by selecting high-quality initial parameters, rather than initializing randomly. These methods can be broadly divided into heuristic and informative approaches. Heuristic initializers are implemented entirely classically and do not require access to a quantum processor; a common example is small-angle or ``identity'' initialization, where parameters are set near zero so that the initial circuit closely approximates the identity operation \cite{zhang2022gaussian, park2024hamiltonian, wang2024trainability, park2024hardware, shi2024avoiding}.  In contrast, informative initializers leverage either prior quantum data or classical surrogate models to construct informed initial parameters. These include warm-start methods~\cite{puig2025variational, mhiri2025unifying}, parameter transfer across related problem instances~\cite{mele2022avoiding,liu2023mitigating,peng2025titan}, and pre-training strategies that use classical emulators, such as Lie-algebraic surrogates~\cite{goh2023lie}, matrix product state–based models~\cite{dborin2022matrix,rudolph2023synergistic}, and neural-network approaches~\cite{ cervera2021meta, shaffer2023surrogate}, to perform substantial portions of the optimization classically before any quantum execution. By reducing the distance to a high-quality solution in parameter space, such initialization schemes significantly lower the quantum sampling cost of subsequent optimization.

\noindent \textit{Intelligent optimizers}.
Intelligent optimizers employ classical machine learning to reduce quantum sampling costs in variational algorithms by predicting optimization trajectories. For instance, meta-learning with recurrent neural networks can generate informed parameter initializations, cutting down the number of quantum-classical iterations needed for convergence~\cite{verdon2019learning}. Other approaches include QuACK, which applies Koopman operator theory to create a linear representation of gradient-based optimization, enabling faster convergence in tasks such as quantum chemistry, and PALQO, which uses physics-informed neural networks to model VQA training and predict multi-step parameter updates from limited quantum data~\cite{luo2024quack,huang2025palqo}. These methods collectively lower sampling overhead by shifting significant parts of the optimization process to classical computation, while maintaining comparable solution quality.

\noindent \textit{Measurement grouping strategies}. Measurement grouping strategies reduce sampling overhead by exploiting compatibility among terms in the target observable (e.g., a Hamiltonian) to jointly estimate multiple terms in a single measurement setting. The central idea is to partition the observable into subsets of mutually commuting or simultaneously measurable operators, thereby minimizing the total number of distinct quantum measurements required. This approach has been widely adopted as a standard technique for shot-efficient expectation estimation in variational algorithms~\cite{2019HugginsNoiseResilientMeasurements,2024ReggioFastpartitioning,2019GokhaleMinimizing}.

The proposed S-ZNE is fully compatible with these broader strategies. We leave their integration for future investigation.

\subsection{Extrapolation functions used in ZNE}\label{app:extrap_method}

The choice of extrapolation function $ g(\cdot) $ significantly influences the performance of ZNE \cite{Cai2023QEM}. Following the notation in the main text, for a given classical input $ \bm{x} $ and observable $ O $, let $ \bm{z} = \{z_1, \dots, z_u\} $ be the vector of noisy expectation value estimates corresponding to noise scales $ \{\lambda_j\}_{j=1}^u $, where each $ z_j $ is either an experimental estimate $ \hat{f}(\bm{x}, O, \lambda_j) $ or a surrogate prediction $ h(\bm{x}, O, \lambda_j) $. The extrapolation function $ g(\cdot) $ maps $ \bm{z} $ to an estimate of the zero-noise expectation $ f(\bm{x}, O) $. Below we detail the functional forms of $ g(\cdot) $ used in our numerical simulations.

\medskip
\noindent\textbf{Linear extrapolation}.  
This method assumes an approximately linear dependence of the observable on the noise scale. For $ u \geq 2 $, the data points $ \{(\lambda_j, z_j)\}_{j=1}^u $ are used to fit a linear model via ordinary least squares, i.e.,
\[
    \arg\min_{a_0, a_1} \sum_{j=1}^u \bigl(z_j - a_0 - a_1 \lambda_j\bigr)^2.
\]
The closed-form solution is given by
\[
    \begin{bmatrix} a_0 \\ a_1 \end{bmatrix}
    = \bigl(V^\top V\bigr)^{-1} V^\top \bm{z},
    \quad
    V = \begin{bmatrix} 1 & \lambda_1 \\ \vdots & \vdots \\ 1 & \lambda_u \end{bmatrix}.
\]
The zero-noise estimate is then obtained by evaluating the fitted model at $ \lambda_0 = 0 $, yielding $ g(\bm{z}) = a_0 $.

\medskip
\noindent\textbf{Quadratic extrapolation}.  
To capture possible nonlinear behavior, this method fits a quadratic model when $ u \geq 3 $, i.e.,
\[
 (b_0, b_1, b_2) = \arg\min_{b_0,b_1,b_2} \sum_{j=1}^u \bigl(z_j - b_0 - b_1 \lambda_j - b_2 \lambda_j^2\bigr)^2.
\]
The solution is
\[
    \begin{bmatrix} b_0 \\ b_1 \\ b_2 \end{bmatrix}
    = \bigl(V_2^\top V_2\bigr)^{-1} V_2^\top \bm{z},
    \quad
    V_2 = \begin{bmatrix} 1 & \lambda_1 & \lambda_1^2 \\ \vdots & \vdots & \vdots \\ 1 & \lambda_u & \lambda_u^2 \end{bmatrix}.
\]
The zero-noise estimate is again the constant term with $ g(\bm{z}) = b_0 $.

\medskip
\noindent\textbf{Richardson extrapolation}.  
This approach assumes the underlying noise dependence can be modeled by a polynomial of degree at most $ u-1 $. The zero-noise limit is obtained by constructing the unique polynomial of degree at most $ u-1 $ that interpolates the $ u $ points $ \{(\lambda_j, z_j)\}_{j=1} ^u$, and evaluating it at $ \lambda_0 = 0 $. Mathematically, the extrapolated value is given by the Lagrange interpolation formula with
\[
g(\bm{z}) = \sum_{j=1}^u \gamma_j z_j,
\quad
\gamma_j = \prod_{\substack{k=1 \\ k \neq j}}^u \frac{\lambda_k}{\lambda_k - \lambda_j}.
\]
This yields the exact zero-noise value if the observable varies polynomially with $ \lambda $ of degree less than $ u $.

\section{Proof of Theorem 1}\label{apd:proof_theorem2}

In this section, we analyze the total estimation error of S-ZNE, comparing it to that of conventional ZNE. Following notations in the main text, we denote $f(\bm{x}, O) \equiv f(\bm{x}, O, \lambda_0)=\Tr(\rho(\bm{x})O)$ as the ideal zero-noise expectation value. Let $g(\cdot)$ be the extrapolation function mapping a vector of noisy expectation values at different noise levels to an estimate of the zero-noise value. 

Recall that the three types of data vectors used in ZNE or S-ZNE for a given input $\bm{x}$, i.e.,
\begin{itemize}
    \item $\bm{z}_I(\bm{x}) = \{f(\bm{x},O,\lambda_1),\cdots,f(\bm{x},O,\lambda_u)\}$: The ideal vector of exact expectation values at with noise levels $\{\lambda_j\}_{j=1}^u$.
    \item $\bm{z}_S(\bm{x}) = \{h_{\mathsf{qs}}(\bm{x},O,\lambda_1),\cdots,h_{\mathsf{qs}}(\bm{x},O,\lambda_u)\}$: The vector obtained using the classical surrogate predictions $h_{\mathsf{qs}}(\bm{x},O,\lambda_j)$ for S-ZNE with noise levels $\{\lambda_j\}_{j=1}^u$.
    \item $\bm{z}_C(\bm{x}) = \{\hat{f}(\bm{x},O,\lambda_1),\cdots,\hat{f}(\bm{x},O,\lambda_u)\}$: The vector obtained using experimental estimates $\hat{f}(\bm{x},O,\lambda_j)$ from $M$ measurements for conventional ZNE with noise levels $\{\lambda_j\}_{j=1}^u$.
\end{itemize}

Our goal is to derive the upper bound for the average performance between S-ZNE and noiseless expectation values in terms of MSE, i.e., $\mathbb{E}_{\bm{x}\sim\mathbb{D}}|f(\bm{x},O)-g(\bm{z}_S(\bm{x}))|^2$, as well as the average performance between conventional ZNE and noiseless expectation values, i.e., $\mathbb{E}_{\bm{x}\sim\mathbb{D}}|f(\bm{x},O)-g(\bm{z}_C(\bm{x}))|$. The expectation is taken over an arbitrary distribution $\mathbb{D}$ supported on the interval $[-R, R]$. To achieve this goal, we leverage the result of Lemma~\ref{lemma:surrogate_error_correlated}.

\medskip
Now we are ready to present the formal statement of Theorem 1 and the proof details.

\begin{theorem-non}[Formal statement of Theorem 1] Suppose the explored family of circuits $U(\bm{x})$ undergoes Pauli noise channel in Eq.~(\ref{append:eqn:Pauli-channel}) and has correlated parameters $\bm{x} \in [-R,R]^d$ sampled from a distribution $\mathbb{D}$. 
Let $f(\bm{x}, O)$ be the ideal zero-noise limit and $g(\cdot)$ be the employed extrapolation function with the Lipschitz constant $L$. Denote $\zeta^2 = \mathbb{E}_{\bm{x}\sim\mathbb{D}}|f(\bm{x},O) - g(\bm{z}_I(\bm{x}))|^2$ as the intrinsic extrapolation error with $\bm{z}_I(\bm{x})=\{f(\bx, O, \lambda_1), ..., f(\bx, O, \lambda_u)\}$, where $\mathbb{D}$ is an arbitrary distribution supported on the interval $[-R, R]$. When the number of measurements adopted by the conventional ZNE is fixed to be $M$, with probability at least $1-0.05u$, its average performance is upper-bounded by  
\begin{equation}
\mathbb{E}_{\bm{x}}|f(\bm{x},O)-g(\bm{z}_C(\bm{x}))|^2 \le \zeta^2 + \frac{4L^2uB^2}{M}\ln(40). \nonumber
\end{equation}

Following notations in Lemma~\ref{lemma:surrogate_error_correlated}, when (i) $\epsilon_l\le\sqrt{\epsilon}/4$, (ii) the frequency threshold satisfies  $\Lambda>deq(1+R)$, and (iii) the number of training examples for each regression-based surrogate satisfies 
\begin{equation}
    n_j \geq \frac{64B^2M^2}{3} \left(\frac{de}{\Lambda}\right)^{4\Lambda} \cdot\frac{\log(1/\delta)}{9},
\end{equation}
with probability at least $1-u\delta$, the average performance of S-ZNE is upper bounded by
\begin{equation}
\mathbb{E}_{\bm{x}}|f(\bm{x},O)-g(\bm{z}_S(\bm{x}))|^2 \le \zeta^2 + \frac{4L^2uB^2}{M}\ln(40). \nonumber 
\end{equation}
\end{theorem-non}

\begin{proof}[Proof of Theorem 1]
We first analyze the upper bound of MSE between the ideal results and the outputs of the conventional ZNE. To achieve this goal, we leverage the extrapolated values under the ideal setting, i.e., $g(\bm{z}_I(\bm{x}))$. Accordingly, the corresponding upper bound is
\begin{align}
\mathbb{E}_{\bm{x}\sim\mathbb{D}}|f(\bm{x},O)-g(\bm{z}_S(\bm{x}))|^2  &= \mathbb{E}_{\bm{x}\sim\mathbb{D}}|f(\bm{x},O) - g(\bm{z}_I(\bm{x})) + g(\bm{z}_I(\bm{x})) - g(\bm{z}_S(\bm{x}))|^2 \\
         &\leq 2\mathbb{E}_{\bm{x}\sim\mathbb{D}}|f(\bm{x},O) - g(\bm{z}_I(\bm{x}))|^2 + 2\mathbb{E}_{\bm{x}\sim\mathbb{D}}|g(\bm{z}_I(\bm{x})) - g(\bm{z}_S(\bm{x}))|^2  \\
         & = 2\zeta^2 + 2\mathbb{E}_{\bm{x}\sim\mathbb{D}}|g(\bm{z}_I(\bm{x})) - g(\bm{z}_S(\bm{x}))|^2 \label{eq:error_decomp_szne}
    \end{align}
where the inequality comes from the triangle inequality, and the last equality follows the definition of $\zeta^2$.

Similarly, for S-ZNE, we can apply the same decomposition strategy to obtain the upper bound of MSE between S-ZNE and the zero-noise limit is 
\begin{equation}\label{eq:error_decomp_zne}
    \mathbb{E}_{\bm{x}\sim \mathbb{D}}|f(\bm{x},O)-g(\bm{z}_C(\bm{x}))|^2 \le 2\zeta^2 + 2\mathbb{E}_{\bm{x}\sim \mathbb{D}}|g(\bm{z}_I(\bm{x})) - g(\bm{z}_S(\bm{x}))|^2.
\end{equation}

For both ZNE and S-ZNE, the term $2\zeta^2$ refers to the intrinsic error induced by the selected extrapolation function. In this regard, the second term in Eqs.~(\ref{eq:error_decomp_szne}) and (\ref{eq:error_decomp_zne}) quantifies the error introduced by using finite measurements (i.e., $\bz_C(\bx)$) or the output of the learning surrogate (i.e.,  $\bm{z}_S(\bm{x})$) instead of the ideal values $\bm{z}_I(\bm{x})$. In what follows, we separately derive the upper bound of these two terms.

\medskip
\noindent\underline{Upper bound of the second term}. By exploiting the assumption that the extrapolation function $g$ is Lipschitz continuous with the constant $L$ (with respect to the $\ell_2$ norm). The upper bound of the second term in Eq.~(\ref{eq:error_decomp_zne}) is 
\begin{equation}
    2\mathbb{E}_{\bm{x}}|g(\bm{z}_I(\bm{x}))-g(\bm{z}_C(\bm{x}))|^2 \le 2L^2 \sum_{j=1}^u \mathbb{E}_{\bm{x}}|f(\bm{x},O,\lambda_j)-\hat{f}(\bm{x},O,\lambda_j)|^2. \label{eq:lipschitz_bound_zne}
\end{equation}
Recall that the difference between $f(\bm{x},O,\lambda_j)$ and $\hat{f}(\bm{x},O,\lambda_j)$ is caused by the finite $M$ measurements. Supported by the Hoeffding inequality, with probability at least $0.95$, when the number of measurements is $M$, the conventional ZNE at each noise rate $j$ satisfies 
\begin{equation}
    \left|f(\bm{x},O,\lambda_j)-\hat{f}(\bm{x},O,\lambda_j)\right| \le \sqrt{\frac{2B^2}{M}\ln(\frac{2}{0.05})}. \label{eq:shot_noise_term_bound}
\end{equation}
Combining the above two inequalities, the upper bound in Eq.~(\ref{eq:error_decomp_zne}) yields
\begin{equation}\label{eq:shot_noise_term_bound_avg}
2\mathbb{E}_{\bm{x}\sim \mathbb{D}}|g(\bm{z}_I(\bm{x}))-g(\bm{z}_C(\bm{x}))|^2 \le 2L^2  \frac{2B^2}{M}\ln(\frac{2}{0.05})  =  {\frac{4L^2uB^2}{M}\ln(40)}.
\end{equation}

\smallskip
We next derive the upper bound of the second term in Eq.~(\ref{eq:error_decomp_szne}), which amounts to the MSE between the ideal extrapolation and the extrapolation by the outputs from the regression-based learning surrogates $\{h_{\mathsf{qs}}(\bx, O, \lambda_j)\}$. As with the conventional ZNE, the property of Lipschitz continuity of $g(\cdot)$ gives
\begin{equation}
2\mathbb{E}_{\bm{x}\sim \mathbb{D}}|g(\bm{z}_I(\bm{x}))-g(\bm{z}_S(\bm{x}))|^2 \le 2L^2 \mathbb{E}_{\bm{x}\sim \mathbb{D}}\|\bm{z}_I(\bm{x}) - \bm{z}_S(\bm{x})\|_2^2 \leq 2L^2 \sum_{j=1}^u \mathbb{E}_{\bm{x}\sim \mathbb{D}}|f(\bm{x},O,\lambda_j)-h_{\mathsf{qs}}(\bm{x},O,\lambda_j)|^2, \label{eq:lipschitz_bound_szne}
\end{equation}
where the last inequality employs the triangular inequality. 

To attain a comparable performance with conventional ZNE, it amounts to analyzing the required number of training examples $n_j$ to ensure that at each noise level $\lambda_j$, the prediction error of the regression-based surrogate, i.e., $\epsilon_j := \mathbb{E}_{\bx\sim \mathbb{D}}|f(\bm{x},O,\lambda_j)-h_{\mathsf{qs}}(\bm{x}, O,\lambda_j)|^2$,  is well bounded. More specifically, according to Eq.~(\ref{eq:shot_noise_term_bound}), such error should be bounded by ${2B^2}\ln(40)/M$. By substituting this quantity into Lemma~\ref{lemma:surrogate_error_correlated}, we need to derive the required number of training examples $n_j$ such that the  average prediction error satisfies  
\[\epsilon_j = 16B^2(deq(1+R)/\Lambda)^{2\Lambda} \leq \frac{2B^2\ln(40)}{M}.  \]

To achieve this goal, we first reformulate the required number of training examples $n_j$ of Lemma~\ref{lemma:surrogate_error_correlated} in terms of $\epsilon_j$. Formally, with probability at least $1-\delta$, the number of training examples yields
\begin{align}
    n_j & = \left(\frac{1}{q(1+R)}\right)^{4\Lambda}\cdot\frac{\log(1/\delta)}{9} \nonumber\\
    & = \left(\frac{de}{\Lambda}\right)^{4\Lambda}\left(\frac{\Lambda}{deq(1+R)}\right)^{4\Lambda}\cdot\frac{\log(1/\delta)}{9} \\
    & =   \left(\frac{de}{\Lambda}\right)^{4\Lambda} \left(\frac{deq(1+R)}{\Lambda} \right)^{-4\Lambda}\cdot\frac{\log(1/\delta)}{9} \\
    & =   \left(\frac{de}{\Lambda}\right)^{4\Lambda} \Big[\frac{16B^2}{16B^2}\left(\frac{deq(1+R)}{\Lambda} \right)^{2\Lambda}\Big]^{-2}\cdot\frac{\log(1/\delta)}{9} \\
    & =  256B^4  \left(\frac{de}{\Lambda}\right)^{4\Lambda} \frac{1}{\epsilon_j^2}\cdot\frac{\log(1/\delta)}{9}.
\end{align}

When the error threshold is ${2B^2\ln(40)}/{M}$, with probability at least $1-\delta$, the corresponding number of training examples is  
\[n_j \geq 256B^4  \left(\frac{de}{\Lambda}\right)^{4\Lambda} \frac{M^2}{12B^2}\cdot\frac{\log(1/\delta)}{9} = \frac{64B^2M^2}{3} \left(\frac{de}{\Lambda}\right)^{4\Lambda} \cdot\frac{\log(1/\delta)}{9}.\]

Supported by the union bound, when the number of training examples $n_j$ used in each regression-based learning surrogate exceeds the above threshold, with probability at least $1-u\delta$, we have 
\begin{equation} 
\mathbb{E}_{\bm{x}\sim \mathbb{D}}|f(\bm{x},O)-g(\bm{z}_C(\bm{x}))|^2 \le 2\zeta^2 + {\frac{4L^2uB^2}{M}\ln(40)}.
\end{equation}

\end{proof}

\subsection{The Lipschitz constant of the extrapolation function}

Theorem~1 indicates that the performance of both conventional ZNEs and S-ZNE depends on the Lipschitz constant $L$. In this subsection, we comprehend the scaling of $L$ when the linear extrapolation function $g(\cdot)$ introduced in SI~\ref{app:extrap_method} is exploited. In addition, the unitary folding is adopted to construct the vectors $\bz_I$, $\bz_C$, and $\bz_S$. Without loss of generality, we use $\bz$ to denote these three vectors, where each entry $ z_j $ is either an ideal result $f(\bx, O, \lambda_j)$, an experimental estimation $\hat{f}(\bm{x}, O, \lambda_j)$, or a surrogate prediction $ h(\bm{x}, O, \lambda_j) $.

When the linear extrapolation function is employed, its output is obtained via least-squares regression. Mathematically, we have
\begin{equation}\label{append:eqn:Lip}
    g(\bz(\bx)) = a_0 = \langle \bm{s}, \bz \rangle\quad \text{with} \quad \bm{s}^{\top}=[1,0] (W^{\top}W)^{-1}W^{\top},
\end{equation}
where $W= \big [ \begin{smallmatrix} 
1, & 1, & \cdots, & 1 \\
1, & 2, & \cdots, & u
\end{smallmatrix} \big]^{\top}$. In this regard, the Lipschitz constant $L$ can be derived by analyzing the upper bounded of the $\ell_2$ norm of $\bm{s}$, i.e.,
\[ L = \sqrt{\sum_{i=1}^u |\bm{s}_i|^2}. \]

In what follows, we derive the analytical form of each entry in $\bm{s}$. Specifically, the matrix $W^{\top}W \in \mathbb{R}^{2\times 2}$ equals to 
\[ W^{\top}W = \Big[\begin{matrix}
    u & \sum_{j=1}^u j \\
    \sum_{j=1}^u j & \sum_{j=1}^u j 
\end{matrix} \Big]
= \Big[\begin{matrix}
    u & \frac{u(u+1)}{2} \\
    \frac{u(u+1)}{2} & \frac{u(u+1)(2u+1)}{6}
\end{matrix} \Big]. \] 
According, its inversion equals to 
\[ \left(W^{\top}W\right)^{-1} = \frac{12}{u^2(u^2-1)}\Big[\begin{matrix}
     \frac{u(u+1)(2u+1)}{6} & -\frac{u(u+1)}{2} \\
    -\frac{u(u+1)}{2} & u
\end{matrix} \Big]. \]
Combining this result with Eq.~(\ref{append:eqn:Lip}), the explicit form of the $i$-th entry of $\bm{s}$ is
\begin{equation}
    \bm{s}_i = \frac{12}{u^2(u^2-1)}\left(\frac{u(u+1)(2u+1)}{6} - \frac{u(u+1)}{2}i \right) = \frac{2u+1-6i}{u(u-1)}. 
\end{equation}
Thereby, we have
\begin{equation}
    L = \sqrt{\sum_{i=1}^u \left|\frac{2u+1-6i}{u(u-1)}\right|^2} =  \sqrt{\frac{(2u+1)^2}{u(u-1)^2}}.
\end{equation}
In this regard, we can conclude that the Lipshitz constant $L$ monotonically decreases with increasing $u$. When $u\rightarrow \infty$, we have $L\rightarrow \sqrt{4/u}$. 
 
\section{Extension of S-ZNE in the hybrid scenario}\label{append:sec:h-s-ZNE}

While the main text focuses on a complete substitution of quantum circuit executions with classical surrogate predictions within the ZNE framework (termed S-ZNE), we present here an extension involving {partial substitution}, yielding a hybrid paradigm for extrapolation. An intuition is illustrated in Fig.~\ref{fig:scheme}. Compared to the original S-ZNE, this variant offers a flexible trade-off between the potential accuracy gains from retaining direct quantum measurements at low noise levels and the substantial resource savings afforded by surrogates, particularly at high noise levels where surrogate modeling is often more effective.

\medskip
\noindent{\textbf{Methodology}}. The initial stages of this hybrid approach precisely mirror the full S-ZNE protocol described in the main text. The implementations of the first two steps are summarized below.

\begin{enumerate}
    \item {Data Acquisition}. A training dataset $\mathcal{T}(\lambda_j)=\{\bm{x}^{(i,j)},y^{(i,j)}\}_{i=1}^{n_j}$ is collected via quantum measurements (e.g., using classical shadows or direct expectation value estimation) for parameter samples $\bm{x}^{(i)}$ across all $u$ noise levels $\{\lambda_j\}_{j=1}^u$ generated via noise scaling (e.g., unitary folding).

    \item {Surrogate Training}. Based on $\mathcal{T}(\lambda_j)$, $u$ distinct classical learning surrogates, $\{h(\bm{x}, O, \lambda_j)\}_{j=1}^u$, are trained. The surrogate model can be either kernel-based (as $h_{\mathsf{cs}}$ in Eq.~(\ref{append:eqn:h-cs})) or regression-based (as $h_{\mathsf{qs}}$ in Eq.~(\ref{append:eqn:h-qs})), where the training methodology is summarized in SI~\ref{apd:surrogate}.
\end{enumerate}

The departure from the full S-ZNE method occurs in the subsequent \textit{validation and selection stage}.

\begin{enumerate}
    \setcounter{enumi}{2}  
    \item {Validation and Thresholding}. A separate and small validation set of parameter vectors, $\mathcal{X}_{\text{val}}$, is utilized. For each noise level $\lambda_j$, we evaluate the Mean Squared Error (MSE) between the surrogate's predictions and direct quantum circuit executions (obtained via additional quantum measurements {only} for the validation set):
    \begin{equation}
        \text{MSE}(\lambda_j) = \frac{1}{|\mathcal{X}_{\text{val}}|}  \left| h(\bm{x}, O, \lambda_j) - \hat{f}(\bm{x}, O, \lambda_j) \right|^2
    \end{equation}
    Here, $\hat{f}(\bm{x}, O, \lambda_j)$ represents the expectation value obtained from executing the circuit on the quantum processor at noise level $\lambda_j$. We consistently observe that this MSE {tends to decrease as the noise level $\lambda_j$ increases}. This trend is warranted by Lemma~\ref{lemma:surrogate_error} and Lemma~\ref{lemma:surrogate_error_correlated}. As such, we establish an MSE threshold, $\eta$. After that, the set of noise levels (typically the highest ones) is identified for which the surrogate model meets this accuracy criterion: $\mathcal{J}_{S} = \{j \mid \text{MSE}(\lambda_j) \le \eta\}$. Let $v = |\mathcal{J}_S|$ be the number of noise levels satisfying this condition.

    \item {Hybrid S-ZNE Construction}. For any given inputs $\bm{x}$ during the {inference} stage, we construct a {hybrid data vector} $\bm{z}_H(\bm{x})$ for extrapolation. This vector selectively combines direct quantum measurements with surrogate predictions, i.e.,
    \begin{equation}
        \bz_H(\bm{x}) = \{ z_j(\bm{x}) \}_{j=1}^u, \quad \text{where} \quad z_j(\bm{x}) = \begin{cases} \hat{f}(\bm{x}, O, \lambda_j) & \text{if } j \notin \mathcal{J}_s \\ h(\bm{x}, O, \lambda_j) & \text{if } j \in \mathcal{J}_s \end{cases}
    \end{equation}

    \item {Extrapolation}. Finally, the same extrapolation functions (e.g., polynomial, Richardson) employed in conventional ZNE and full S-ZNE are applied to the hybrid dataset $\bm{z}_H(\bm{x})$ to estimate the zero-noise expectation value $f(\bm{x}, O, \lambda=0)$.
\end{enumerate}

\noindent{\textbf{Discussion of trade-offs}}. The primary motivation for this hybrid protocol stems from scenarios where classical surrogates might exhibit non-negligible prediction errors, particularly at low noise levels. In such cases, completely replacing $\hat{f}(\bm{x}, O, \lambda_j)$ with $h(\bm{x}, O, \lambda_j)$ could potentially degrade the final extrapolation accuracy compared to conventional ZNE. 

The proposed hybrid approach mitigates this risk by retaining direct quantum measurements for those low-noise data points where the surrogate's fidelity might be lower (i.e., where $\text{MSE}(\lambda_j) > \eta$), while still leveraging the efficiency of surrogates for the higher noise levels where they perform well and where quantum execution (requiring deeper circuits via folding) is most resource-intensive.

However, this potential accuracy retention comes at the cost of {reduced quantum measurement savings} during inference compared to the full S-ZNE approach. For each new parameter vector $\bm{x}$ evaluated, the hybrid method still requires executing the quantum circuit at $u-v$ noise levels. Consequently, the reduction in quantum measurement overhead compared to conventional ZNE is scaled by a factor of $v/u$, determined by the number of noise levels $v$ where the surrogate meets the accuracy threshold $\eta$ set during validation. The choice of $\eta$ thus directly controls the balance between potential accuracy preservation and computational resource savings. Refer to SI~\ref{append:hybrid} for more simulation results.

\section{Additional experimental results of S-ZNE}
\label{app:additional_results}

This section presents more implementation details and simulation results omitted in the main text. In particular, SI~\ref{app:circuits} provides the circuit implementation details for the explored tasks, SI~\ref{app:additional_qaoa} evaluates the robustness of S-ZNE across different extrapolation models, and SI~\ref{app:additional_metrology} investigates its data efficiency in quantum metrology. 

\subsection{Circuit implementation}
\label{app:circuits}

\begin{figure}[htbp]
    \centering
    \includegraphics[width=0.6\textwidth]{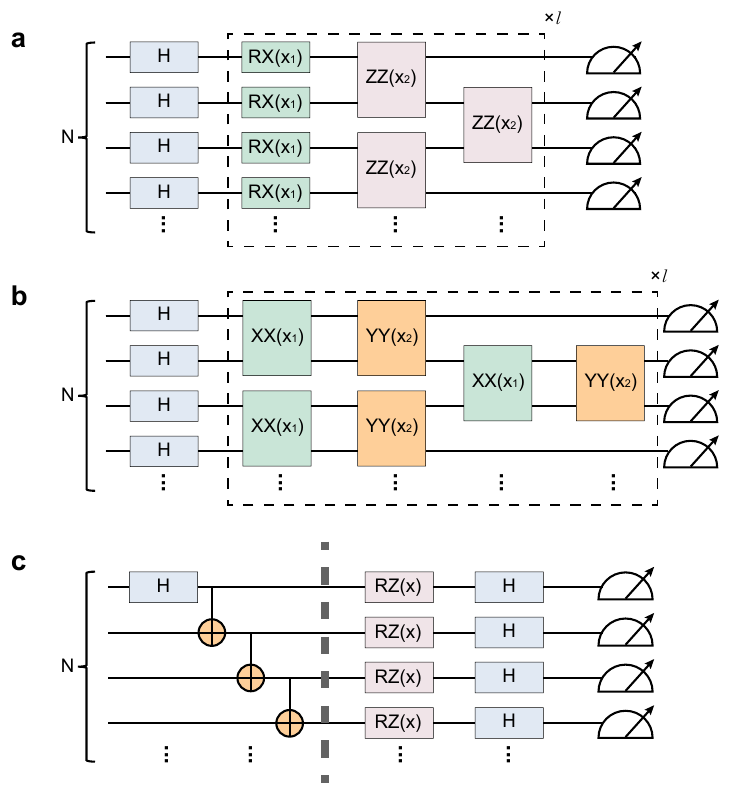}
    \caption{\small\textbf{Ansatz circuits used in numerical experiments.} \textbf{a.} Trotterized Hamiltonian variational ansatz for the 1D Transverse Field Ising Model (TFIM). \textbf{b.}  Trotterized Hamiltonian variational ansatz for the 1D Heisenberg Model (HM). \textbf{c.} Circuit implementation for GHZ-state-based quantum metrology.}
    \label{fig:circuit}
\end{figure}

For the ground state energy estimation tasks, we employ Hamiltonian Variational Ansatz (HVAs) derived from a first-order Trotter-Suzuki decomposition~\cite{lloyd1996universal, stanisic2022observing} of the respective problem Hamiltonians, $\mathrm{H}_{\text{Ising}}$ and $\mathrm{H}_{\text{Heisen}}$. For both Hamiltonians, the ansatz is applied to the uniform superposition state, prepared by applying an initial layer of Hadamard gates $H^{\otimes N}$ to the all-zero state $\ket{0}^{\otimes N}$.

The circuit structure for the 1D TFIM is illustrated in Fig.~\ref{fig:circuit}a. The variational ansatz $U(\bm{x})$ is defined by the unitary:
\begin{equation}
    U(\bm{x}) = \Big[\exp\!\Big(-\imath \bx_1 \sum_{\langle i,j\rangle \in E} Z_i Z_j\Big) \exp\!\Big(-\imath\bx_2 \sum_i X_i\Big)\Big]^l H^{\otimes N}.
\end{equation}
This represents the Trotterized evolution under the Ising Hamiltonian terms. In all simulations, we set the number of layers $l=1$, as this shallow ansatz structure proves sufficient for approximating the ground state energy with high fidelity for the models under study.

The circuit implementation for the 1D Heisenberg model with $J_z = 0$ (the XY model) is visualized in Fig.~\ref{fig:circuit}b. Mathematically, the corresponding unitary operator is given by:
\begin{equation}
    U(\bm{x})=\Big [\exp\!\Big(-\imath \bx_1 \sum_{\langle i,j\rangle \in E}  X_i X_j\Big)  \exp\!\Big(-\imath \bx_2 \sum_{\langle i,j\rangle \in E}  Y_i Y_j\Big)\Big]^l H^{\otimes N}.
\end{equation}
Analogous to the TFIM case, we fix the layer depth to $l=1$ for all simulations.

The circuit implementation for the quantum metrology task is illustrated in Fig.~\ref{fig:circuit}c. This protocol is designed to achieve Heisenberg-limited sensitivity. The circuit first prepares an $N$-qubit GHZ state, $|\mathrm{GHZ}\rangle_N = (|0\rangle^{\otimes N} + |1\rangle^{\otimes N})/\sqrt{2}$, using a Hadamard gate on the first qubit followed by a chain of CNOT gates. Subsequently, the unknown phase $x$ is encoded via a global $\RZ(x)^{\otimes N}$ rotation. This operation imparts a collective relative phase between the two components of the GHZ state:
\begin{equation}
    |\psi(\bx_1)\rangle = \frac{1}{\sqrt{2}} \left( e^{-\imath N x / 2} |0\rangle^{\otimes N} + e^{\imath N x / 2} |1\rangle^{\otimes N} \right),
\end{equation}
resulting in a total relative phase of $e^{\imath N x}$. To convert this phase information into a measurable signal, the expectation value of the global observable $O = X^{\otimes N}$ is measured. Operationally, this measurement is implemented by applying a final layer of Hadamard gates ($H^{\otimes N}$) to all qubits, which rotates the measurement basis, followed by a standard measurement of the global observable $Z^{\otimes N}$. In the noiseless limit, the measured expectation value is $\cos(Nx)$, achieving the Heisenberg-limited phase sensitivity. 

\subsection{Random feature sampling for classical learning surrogates}

In our numerical experiments, the construction of the classical learning surrogates $\{h_\mathsf{qs}\}_{j=1}^u$ employs a randomized feature selection strategy to balance model expressiveness with computational efficiency. Each surrogate is implemented as a linear model using a subsampled feature map, specifically $h(\bx, O, \lambda_j) = \langle \bm{\Phi}_{\Omega(\Lambda)}(\bx), \mathbf{w}_j \rangle$, where $\bm{\Phi}_{\Omega(\Lambda)}(\bx)$ is constructed by randomly and uniformly sampling $n_f$ elements from the complete set of trigonometric monomials $\{\Phi_{\bm{\omega}}(\bx) | \bm{\omega} \in \mathfrak{C}(\Lambda)\}$ with $\mathfrak{C}(\Lambda)=\{\bomega \in \{0,\pm1\}^d \mid \|\bomega\|_0 \le \Lambda\}$.

In our experimental implementation, we set $n_f = 1000$ across all simulations. For ground-state energy estimation tasks, the frequency truncation parameters were set to $\Lambda = 2$ for the transverse field Ising model and $\Lambda = 4$ for the Heisenberg model, while for quantum metrology applications with GHZ states, we used $\Lambda = 2$. These parameter choices were determined through empirical validation to provide an optimal balance between model capacity and generalization performance for their respective problem domains. The randomized feature sampling approach significantly reduces computational overhead while maintaining the theoretical approximation guarantees of the full trigonometric basis, enabling efficient training of surrogates even for high-dimensional parameter spaces.
 
\subsection{Robustness of S-ZNE across varied extrapolation functions}
\label{app:additional_qaoa}

While the main text employs linear extrapolation for its simplicity, we further investigate the robustness of the S-ZNE framework by evaluating its compatibility with a diverse set of extrapolation functions. We benchmark S-ZNE against conventional ZNE under the same setting used for ground-state energy estimation in the transverse-field Ising model (TFIM) and Heisenberg model (HM), as detailed in the main text. For S-ZNE, we use the regression-based classical surrogate $h_{\mathsf{qs}}$ trained with $n_j = 200$ samples per noise level $\lambda_j$; for conventional ZNE, we use the estimate $\hat{f}$ obtained from $M = 1 \times 10^6$ shots per noise level. The only variable in this comparison is the extrapolation function applied to the surrogate data vector $\bm{z}_S(\bm{x})$ and the conventional data vector $\bm{z}_C(\bm{x})$.

We compare three common extrapolation strategies in quantum error mitigation: (i) \textit{Linear} (first-order least-squares regression), (ii) \textit{Quadratic} (second-order least-squares regression), and (iii) \textit{Richardson} extrapolation. Detailed implementations are provided in SI~\ref{app:extrap_method}. To further assess robustness, we perform this comparison under two noise models: globe depolarizing (DP) noise and a composite noise model combining DP with coherent (CO) noise, consistent with the main text.

\begin{figure}[htbp]
    \centering
    \includegraphics[width=0.7\textwidth]{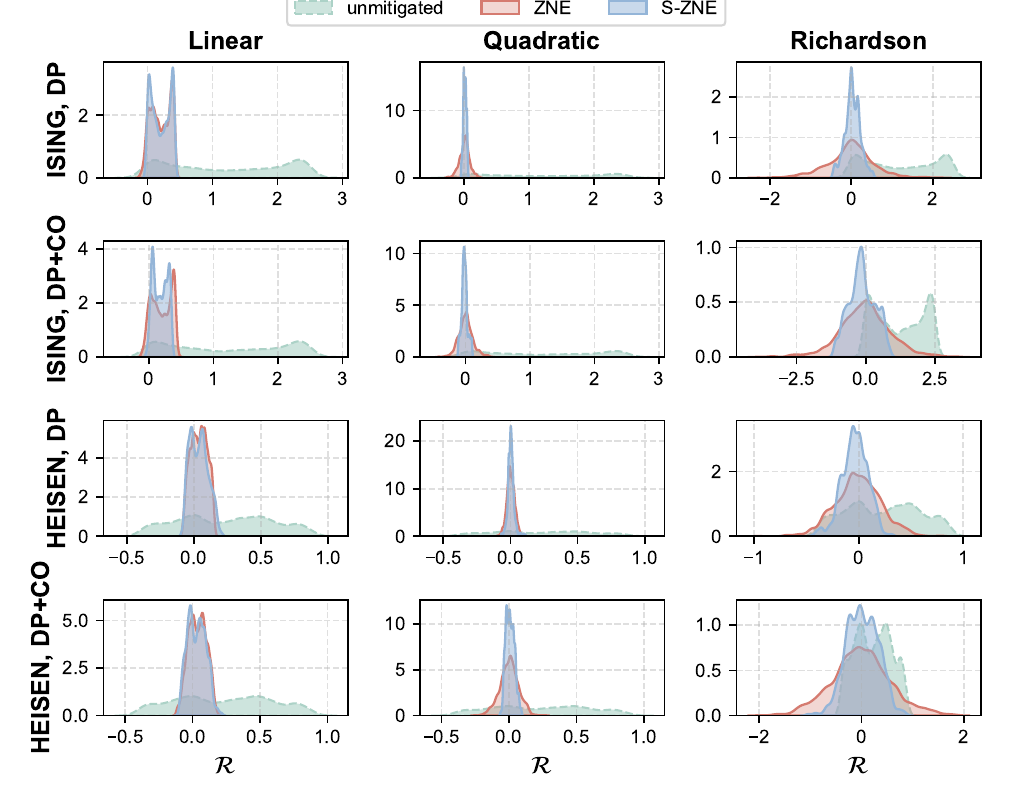}
    \caption{\small\textbf{Residual distributions of conventional ZNE and S-ZNE for different extrapolation functions.} Probability density of residuals (error-mitigate estimation minus ideal value) for unmitigated, ZNE, and S-ZNE results under depolarizing (DP) and DP+coherent (CO) noise. Three extrapolation functions are compared: Linear, Quadratic, and Richardson. Results are aggregated over 1000 test instances.}
    \label{fig:compare_extrap}
\end{figure}

Results are summarized in Fig.~\ref{fig:compare_extrap}, which shows the mitigation residuals for S-ZNE ($\mathcal{R}_S$) and conventional ZNE ($\mathcal{R}_C$). Across all extrapolation functions and both noise models, S-ZNE achieves mitigation accuracy comparable to that of conventional ZNE. Under the mixed DP+CO model, both methods exhibit degraded performance when using Richardson extrapolation, indicating that the surrogate-based approach faithfully preserves the behavior—and limitations—of the underlying extrapolation function without introducing significant additional bias. This confirms that the surrogate can effectively replace direct quantum measurement in the ZNE pipeline, inheriting both the advantages and instabilities of the chosen extrapolation method.

\subsection{Data efficiency in quantum metrology}
\label{app:additional_metrology}
\begin{figure}[htbp]
    \centering
    \includegraphics[width=0.6\textwidth]{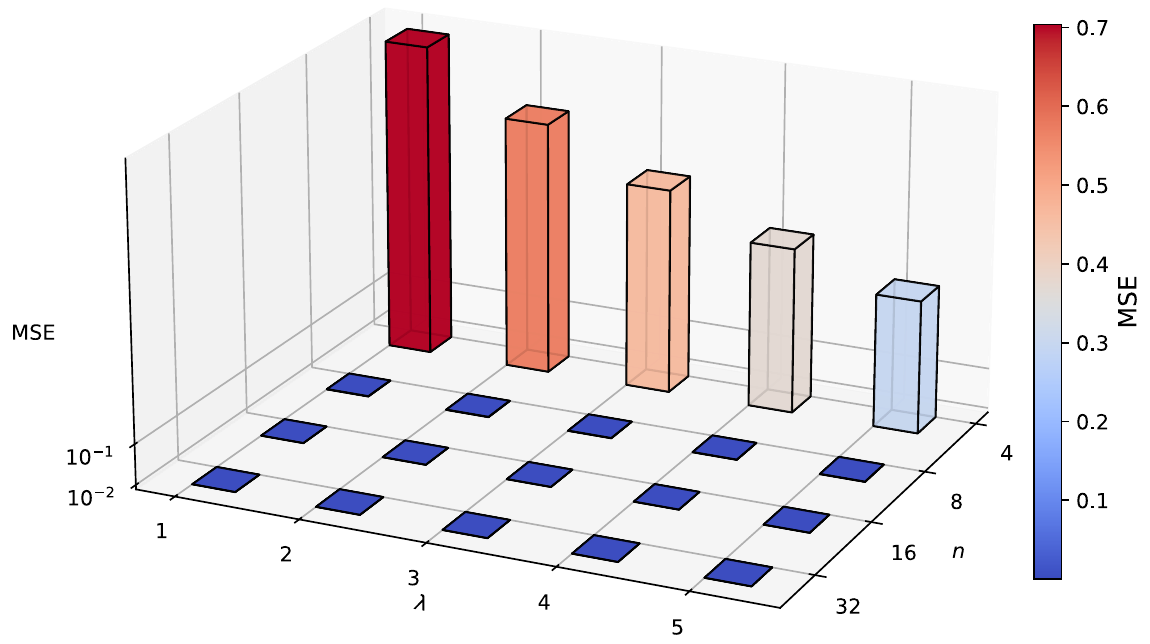}
    \caption{\small\textbf{Surrogate prediction accuracy versus training set size for GHZ metrology.}
    Mean squared error of surrogate predictions as a function of training sample size $n_j \in \{4,8,16,32\}$ and noise factor $\lambda_j$. Results represent averages over 10 independent experiments.}
    \label{fig:prediction_mse}
\end{figure}

To further characterize the data efficiency of S-ZNE in the quantum metrology task, we examine how regression-
based surrogate prediction accuracy depends on the number of training samples $n_j$ per noise level $\lambda_j$. All hyperparameter settings are identical to those introduced in the main text.

Figure~\ref{fig:prediction_mse} plots the mean squared error (MSE) of surrogate predictions as a function of training set size $n_j \in \{4, 8, 16, 32\}$ across $u=5$ noise levels. 
The results demonstrate a sharp, non-linear improvement in accuracy with $n_j$. For a minimal training set of $n_j = 4$, the surrogate exhibits a high prediction error (e.g., MSE $\approx 0.704$ at the $\lambda_j = 1$ noise level), indicating that this sparse dataset is insufficient to construct the surrogate model and capture the phase-dependent response. However, a modest increase to $n_j = 8$ reduces the MSE by over three orders of magnitude (to $\sim {6.17\times 10^{-4}}$ at $\lambda_j=1$). The error continues to decrease rapidly as $n_j$ grows: for $n_j = 16$, the MSE drops to $2.20\times10^{-4}$, and at $n_j = 32$, it reaches $1.60\times10^{-4}$ (all values reported for $\lambda_j=1$). 
This rapid decay in error reveals a distinct threshold effect, suggesting that once a minimal number of training samples is provided to constrain the surrogate's trigonometric feature space, the model generalizes effectively from sparse data.

\section{Simulation results of hybrid S-ZNE}\label{append:hybrid}

The hybrid S-ZNE framework described in SI~\ref{append:sec:h-s-ZNE} addresses a potential limitation of the S-ZNE approach. That is, while surrogates can dramatically reduce quantum measurement costs, their predictive accuracy may be insufficient at low noise levels.  

\begin{figure*}[t!]
    \centering\includegraphics[width=0.9\textwidth]{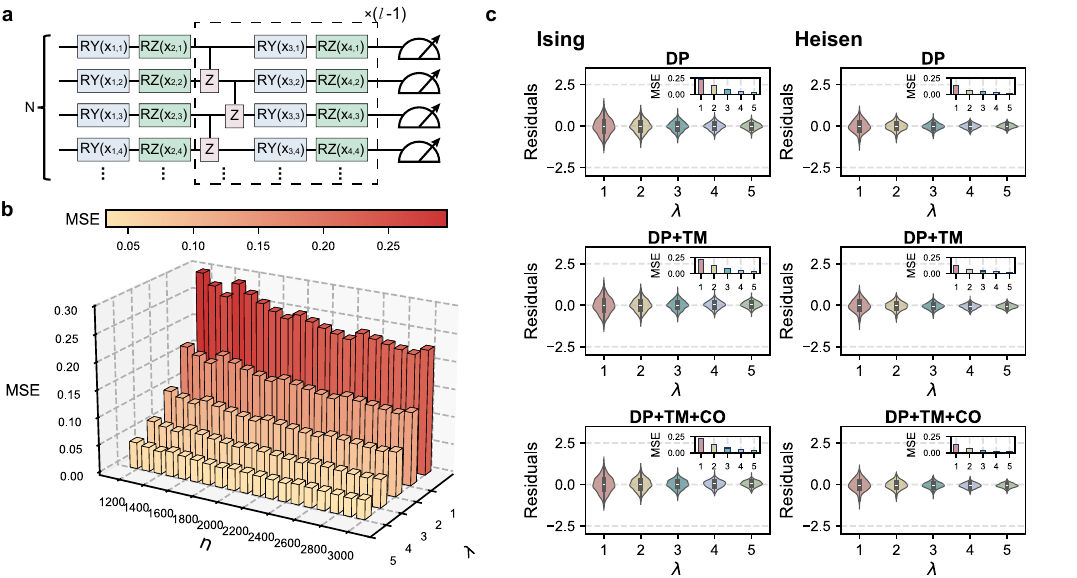}
    \caption{\small\textbf{Circuit architecture and surrogate fidelity across noise models and Hamiltonians.} \textbf{a.} Hardware-efficient ansatz with uncorrelated $\RY$ and $\RZ$ rotations used in simulations. \textbf{b.} Mean squared error (MSE) of surrogate predictions versus training set size $n_j$ and noise scaling factor $\lambda_j$, evaluated on both the 1D transverse-field Ising and Heisenberg models. MSE decreases with larger $n_j$ and higher $\lambda_j$, indicating improved surrogate accuracy in high-noise regimes. \textbf{c.} Residual distributions (prediction minus measurement) for three composite noise models—depolarizing (DP), DP+thermal (TM), and DP+TM+coherent (CO)—shown as violin plots; nested bar charts report corresponding MSE values for both Hamiltonians.}
    \label{fig:veri_diffshadow}
\end{figure*}

To evaluate the performance of hybrid S-ZNE, we perform numerical simulations using a 6-qubit hardware-efficient ansatz, as illustrated in Fig.~\ref{fig:veri_diffshadow}a. We consider both the transverse-field Ising model (TFIM) and the Heisenberg model (HM)(defined in maintext), with a fixed circuit depth of $l = 2$ layers. The model parameters are set as follows: for TFIM, we use a uniform coupling $J = 0.1$ and a transverse field $h = 0.5$; for HM, we set $J_x = 0.1$,$J_y = 0.1$ and $J_z = 0.5$. Noise levels are amplified via unitary folding, and extrapolation is performed over $u = 5$ such levels using linear extrapolation. To assess the robustness of the approach, we incorporate three distinct noise models(detailed in SI~\ref{append:A-3}): local depolarizing (DP) noise, thermal relaxation (TM), and coherent (CO) over-rotation. The corresponding noise parameters are provided in Table~\ref{tab:noise_params}.

 \begin{table}[!h]
\caption{Parameter settings for different noisy channels. Notations follow the definitions in SI~\ref{append:A-3}. }  
\label{tab:noise_params}  
\begin{tabular}{@{}|lll|@{}}
\toprule
\multicolumn{1}{|l|}{\textbf{Noise}} & \multicolumn{1}{l|}{\textbf{Parameter}}                                                    & \textbf{Value}                                                                                                              \\ \midrule
\multicolumn{1}{|l|}{Local depolarizing}    & \multicolumn{1}{l|}{$ p_d$}                                                         & single-qubit gate: $0.001$, two-qubit gate: $0.005$                                                                               \\ \midrule
\multicolumn{1}{|l|}{Thermal}        & \multicolumn{1}{l|}{\begin{tabular}[c]{@{}l@{}}$T_1$\\ $T_2$\\ $t_g$\\ $p_e$\end{tabular}} & \begin{tabular}[c]{@{}l@{}}100000 $us$\\ 30000 $\mu s$\\ single-qubit gate:15 $\mu s$, two-qubit gate: $20$ $\mu s$\\ 0.01\end{tabular} \\ \midrule
\multicolumn{1}{|l|}{Coherent}       & \multicolumn{1}{l|}{$ \theta_P$}                                                             & $\text{Unif}[-0.01\pi , 0.02\pi]$         
\\ \bottomrule
\end{tabular}
\end{table}

We trained the kernel-based surrogates $h_{\mathsf{cs}}$, defined in Eq.~(\ref{append:eqn:h-cs}), with a frequency truncation threshold of $\Lambda = 2$. These surrogates were trained on classical shadow collected at $u = 5$ distinct noise levels with $T = 500$. The validation set $\mathcal{X}_{\rm val}$, which was used to determine the hybrid data vector $\bm{z}_H(\bm{x})$, contained 500 random input points; the ground-truth expectation values for these points were estimated using 40,000 measurement shots each.

To evaluate the data efficiency of the surrogate, we varied the per-noise-level training set size over $n_j \in \{1200, 1400, \dots, 3000\}$. Figure~\ref{fig:veri_diffshadow}b shows that the surrogate's mean squared error (MSE) consistently decreases as $n_j$ increases. Notably, this improvement is more substantial at higher noise levels $\lambda_j$. This trend holds across all tested noise models, including the composite DP+TM+CO model (Fig.~\ref{fig:veri_diffshadow}c). For instance, in the transverse-field Ising model (TFIM) simulation, the MSE drops from approximately 0.23 at $\lambda_j = 1$ to about 0.03 at $\lambda_j = 5$. A similar reduction is observed for the Heisenberg model (HM), where the MSE falls from about 0.13 to roughly 0.02 over the same range of noise levels.

\begin{figure*}[tb]
    \centering\includegraphics[width=0.8\textwidth]{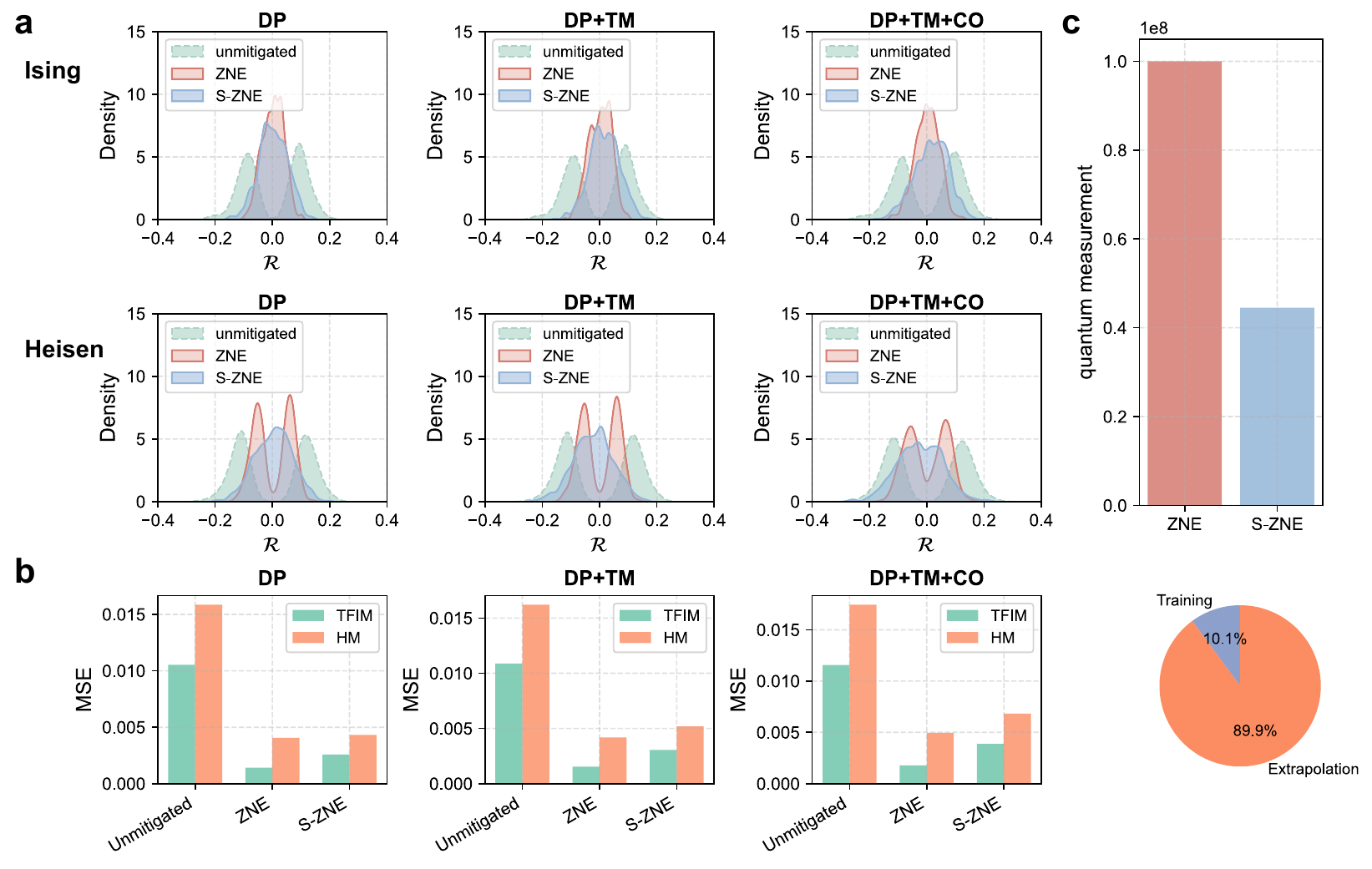}\vspace{-10pt}
    \caption{\small\textbf{Error mitigation performance and quantum resource efficiency for Ising and Heisenberg models.} \textbf{a.} Residual distributions (mitigated estimate minus ideal value) for unmitigated, ZNE, and hybrid S-ZNE results across three noise configurations (DP, DP+TM, and DP+TM+CO), based on 500 test instances with $|f(\bx,O)| > 0.5$. \textbf{b.} Corresponding MSE values confirm that hybrid S-ZNE matches conventional ZNE in accuracy for both Hamiltonians. \textbf{c.} Quantum resource comparison: hybrid S-ZNE incurs a fixed offline training cost and avoids repeated measurements at high $\lambda_j$, reducing per-instance quantum overhead by nearly 60\% compared to conventional ZNE.}
    \label{fig:distribution}
\end{figure*}

\smallskip
Based on this finding, we  define a substitution threshold $\eta = 0.1$ and observe from our validation that the surrogate MSE consistently falls below this threshold only for $\lambda_j \ge 3$. 
We therefore construct a hybrid data vector $\bm{z}_H(\bm{x})$ by retaining direct quantum measurements (using $M = 40,000$ shots) 
for the two lowest noise levels ($\lambda_j = 1, 2$) and using surrogate predictions $h(\bm{x}, O, \lambda_j)$ for the three highest ($\lambda_j = 3, 4, 5$). This hybrid protocol reduces the {per-instance} quantum measurement cost by 60\% compared to conventional ZNE, while retaining the high-fidelity low-noise data crucial for stable extrapolation.

We evaluate the end-to-end mitigation performance on 500 test instances selected to have non-trivial ideal expectation values ($|f(\bm{x},O)| > 0.5$). As defined in the main text, we analyze the mitigation residual $\mathcal{R} = g(\bm{z}) - f(\bm{x},O)$. Fig.~\ref{fig:distribution}a shows the residual distributions for Hybrid S-ZNE ($\mathcal{R}_H$, using $\bm{z}_H$) and conventional ZNE ($\mathcal{R}_C$, using $\bm{z}_C$) under the three noise configurations. Both methods produce residuals tightly centered at zero. The corresponding MSE values reported in Fig.~\ref{fig:distribution}b confirm this quantitatively: Hybrid S-ZNE matches the mitigation accuracy of conventional ZNE, despite foregoing quantum measurements at the three highest (and most costly) noise levels.

Finally, Fig.~\ref{fig:distribution}c highlights the resource trade-off. Conventional ZNE requires $u \times M = 5 \times 40,000 = 200,000$ shots for {each} evaluation. For 500 test samples, this totals $10^8$ measurements. Hybrid S-ZNE, in contrast, requires a {one-time} offline training cost of $n \times u \times T = 3000 \times 5 \times 500 = 7.5\times10^6$ measurements (using $n_j=3000$ for this example). The {per-instance} extrapolation cost is reduced to 40\% of conventional ZNE (retaining $\lambda=1,2$), totaling $0.4 \times (500 \times 200,000) = 4 \times 10^7$ measurements for the 500 samples. The total hybrid S-ZNE cost is thus $7.5\times10^6 + 4\times10^7 = 4.75\times10^7$ measurements, a saving of over 50\%. In this specific task, the one-time training cost accounts for $\approx 16\%$ of the total cost. For applications requiring many repeated evaluations ($N_{\text{eval}} \gg n \times T / ((u-v)M)$), this training cost is amortized, and the cumulative savings asymptotically approach the 60\% per-instance reduction. These results demonstrate that the hybrid S-ZNE framework successfully balances mitigation fidelity with practical resource efficiency.

\end{document}